\documentclass[twocolumn]{aastex701}

\usepackage{multirow}

\usepackage[utf8]{inputenc}
\usepackage{natbib}
\usepackage{graphicx}
\usepackage{longtable}
\usepackage{enumerate}
\usepackage{amsmath}
\usepackage{hyperref}
\usepackage{CJK}
\usepackage[title]{appendix}
\usepackage{rotating}
\usepackage{longtable}

\usepackage{amsmath}

\def\msun{\ifmmode {\rm M}_{\mathord\odot}\else $M_{\mathord\odot}$\fi}
\def\rsun{\ifmmode {\rm R}_{\mathord\odot}\else $R_{\mathord\odot}$\fi}
\def\lsun{\ifmmode {\rm L}_{\mathord\odot}\else $L_{\mathord\odot}$\fi}

\def\co{$^{12}$CO}
\def\c18o{C$^{18}$O}
\def\h2{H$_{2}$}
\def\13co{$^{13}$CO}
\def\n2hp{$_{2}$H$^{+}$}

\def\radmc{{\sc radmc-3d}}
\def\cm2{cm$^{-2}$}

\newcommand{\kms}{km~s$^{-1}$}

\def\casa{{\sc casa}}
\def\deg{$^{\circ}$}

\newcommand{\xd}[1]{\textcolor{black}{{#1}}}
\newcommand{\xdtwo}[1]{\textcolor{black}{{#1}}}

\shorttitle{}
\shortauthors{}

\begin{document}
\begin{CJK*}{UTF8}{gbsn}    

\title{Disk Wind Feedback from High-mass Protostars. V. Application of Multi-Modal Machine Learning to Characterize Outflow Properties}

\author[0000-0001-6216-8931]{Duo Xu}
\affiliation{Canadian Institute for Theoretical Astrophysics, University of Toronto, 60 St. George Street, Toronto, ON M5S 3H8, Canada}
\affiliation{Department of Astronomy, University of Virginia, Charlottesville, VA 22904-4235, USA}
\email[show]{xuduo@cita.utoronto.ca}

\author[0000-0002-4110-8769]{Ioana~A.~Stelea}
\affiliation{Department of Astronomy, Columbia University, New York, NY 10027, USA}
\affiliation{Department of Astronomy, University of Wisconsin-Madison, Madison, WI 53706-1507, USA}
\email{stelea@wisc.edu}

\author[0000-0003-2573-9832]{Joshua S. Speagle (沈佳士)}
\affiliation{Department of Statistical Sciences, University of Toronto, 9th Floor, Ontario Power Building, 700 University Ave, Toronto, ON M5G 1Z5, Canada}
\affiliation{David A. Dunlap Department of Astronomy \& Astrophysics, University of Toronto, 50 St. George Street, Toronto, ON M5S 3H4, Canada}
\affiliation{Dunlap Institute for Astronomy \& Astrophysics, University of Toronto, 50 St. George Street, Toronto, ON M5S 3H4, Canada}
\affiliation{Data Sciences Institute, University of Toronto, 17th Floor, Ontario Power Building, 700 University Ave, Toronto, ON M5G 1Z5, Canada}
\email{}

\author[0000-0001-7511-0034]{Yichen Zhang}
\affiliation{Department of Astronomy, School of Physics and Astronomy, Shanghai Jiao Tong University, 800 Dongchuan Road, Shanghai 200240, China}
\affiliation{State Key Laboratory of Dark Matter Physics, School of Physics and Astronomy, Shanghai Jiao Tong University, Shanghai 200240, China}
\affiliation{Key Laboratory for Particle Astrophysics and Cosmology (MOE) / Shanghai Key Laboratory for Particle Physics and Cosmology, Shanghai 200240, China}
\email{}

\author[0000-0002-3389-9142]{Jonathan C. Tan}
\affiliation{Department of Astronomy and Virginia Institute for Theoretical Astronomy (VITA), University of Virginia, Charlottesville, VA 22904-4235, USA}
\affiliation{Department of Physics \& Astronomy, Chalmers University of Technology, SE-412 96 Gothenburg, Sweden}
\email{}


\begin{abstract}
Characterizing protostellar outflows is fundamental to understanding star formation feedback, yet traditional methods are often hindered by projection effects and complex morphologies. We present a multi-modal deep learning framework that jointly leverages spatial and spectral information from CO observations to infer protostellar mass, inclination, and position angle ($PA$). Our model, trained on synthetic ALMA observations generated from 3D magnetohydrodynamic simulations, utilizes a cross-attention fusion mechanism to integrate morphological and kinematic features with probabilistic uncertainty estimation. Our results demonstrate that Vision Transformer architectures significantly outperform convolutional networks, showing remarkable robustness to reduced spatial resolution. Interpretability analysis reveals a physically consistent hierarchy: spatial features dominate across all parameters, whereas spectral profiles provide secondary constraints for mass and inclination. Applied to observational ALMA data, the framework delivers stable mass and $PA$ estimates with exceptionally tightly constrained inclination angles. This study establishes multi-modal deep learning as a powerful, interpretable tool for overcoming projection biases in high-mass star formation studies.


\end{abstract}

\section{Introduction}
\label{Introduction}

The formation of stars from the gravitational collapse of dense molecular clouds is a fundamental process that drives the evolution of galaxies and provides foundations for planetary systems. During the earliest stages of protostellar evolution, young stars launch bipolar protostellar outflows that play a crucial role in removing angular momentum from the accretion disk, regulating the inflow of material, and injecting mass, momentum, and energy into the surrounding interstellar medium \citep[e.g.,][]{2012A&A...543A.128J,2014prpl.conf..451F,2016ARA&A..54..491B}. This feedback significantly influences both the structure of molecular clouds and the overall efficiency of star formation.

Several theoretical models have been proposed to explain the launching of protostellar outflows, including the X-wind model \citep{1994ApJ...429..781S} and the magnetic tower model \citep{1996MNRAS.279..389L}. Among these, the magnetocentrifugal disk wind model has emerged as the leading paradigm \citep{1982MNRAS.199..883B,1983ApJ...274..677P}. In this model, open magnetic field lines thread the surface layers of a Keplerian accretion disk, allowing gas to be centrifugally accelerated outward, forming a characteristic bipolar outflow composed of a fast, collimated jet and a slower, wide-angle wind. Once launched, this primary disk wind sweeps up and accelerates the surrounding ambient envelope material, producing the massive, entrained outflows that are most commonly observed in molecular tracers such as CO. In fact, it is only in the rare cases where rotating molecular gas is explicitly resolved that the primary disk wind itself is believed to be directly observed. This mechanism has been extensively investigated through sophisticated magnetohydrodynamic (MHD) simulations incorporating diverse physical effects and environmental conditions \citep[e.g.,][]{2003ApJ...582..292O,2015ApJ...801...84G,2020ApJ...900...59M,2022ApJ...941..202R}.


\xdtwo{A growing body of observational evidence, particularly from high-resolution facilities such as ALMA, supports this model in low- and intermediate-mass protostars. Observations reveal outflows with rotating molecular gas and conical wind morphologies, notably in sources such as HH 212 \citep{2021ApJ...907L..41L, 2024ApJ...977..126L}, L1527 \citep{2017MNRAS.467L..76S}, TMC-1A \citep{2016Natur.540..406B}, DG Tau B \citep{2022A&A...668A..78D}, NGC 1333 IRAS 4C \citep{2018ApJ...864...76Z}, and CB 26 \citep{2009A&A...494..147L, 2023A&A...678A.135L}. These findings, which are consistently matched with predictions from MHD disk wind theory, suggest that magnetocentrifugal launching may be a universal mechanism operating across a broad range of stellar masses.} In the high-mass regime, a prominent example of such physics is the rotating bipolar outflow associated with Orion-KL Source I, where high-resolution observations have resolved a clear disk-wind structure \citep{2017NatAs...1E.146H, 2020ApJ...904..158L,2024ApJ...974..150W}. This source provides critical evidence that magnetocentrifugal winds can successfully remove angular momentum and drive powerful outflows even in the presence of the intense radiative feedback characteristic of massive protostars.

Despite these advances, the formation of high-mass stars remains less well understood due to the scarcity of nearby examples, their rapid evolution, and strong radiative feedback. Nevertheless, observations show that massive protostars also launch powerful, collimated outflows resembling scaled-up versions of their low-mass counterparts \citep[e.g.,][]{1996ApJ...472..225S,2004ApJ...608..330B,2019NatCo..10.3630F,2019ApJ...873...73Z,2025ApJ...990..173C}. Among the various theoretical models for massive star formation, i.e., core accretion, competitive accretion, and stellar collisions, the Turbulent Core Accretion (TCA) model \citep{2002Natur.416...59M,2003ApJ...585..850M,2011ApJ...733...55Z,2013ApJ...766...86Z,2014ApJ...788..166Z,2018ApJ...853...18Z} offers a physically motivated extension of low-mass star formation theory, wherein massive, turbulent, magnetized, gravitationally bound cores collapse to form massive stars via disk-mediated accretion \citep[see][for a review]{2014prpl.conf..149T}.

To explore the impact and appearance of protostellar outflows in this model, a series of 3D MHD simulations were carried out by \citet{2019ApJ...882..123S,2023ApJ...947...40S} (hereafter Papers I and II). These simulations follow the collapse of a $60\:M_\odot$ core embedded in a 1~g~cm$^{-2}$ clump, tracking the time evolution of the disk-outflow system as the protostar grows from $\sim1$ to $\sim 24\:M_\odot$. The simulations capture both the structural and kinematic evolution of the magnetocentrifugal disk wind and its interaction with the infalling envelope, offering a rich, physically consistent dataset for comparison with observations.

To bridge theory and observation, Paper III in this series \citep{2024ApJ...966..117X} conducted radiative transfer post-processing of the simulations to produce synthetic CO line emission data, including position-position-velocity (PPV) cubes and spectra. These were further processed with \casa\ to simulate realistic ALMA observations. The resulting synthetic spectra were shown to closely reproduce the features of real ALMA-observed protostellar outflows. By comparing synthetic and observed spectra, Paper III derived best-fit protostellar masses and inclination angles for several real systems. However, the approach was based solely on 1D spectral profiles, without incorporating the rich morphological information present in 2D integrated intensity maps of the red- and blue-shifted outflow lobes. This limitation motivates the development of more comprehensive methods that can simultaneously interpret both the spectral and spatial dimensions of the data.

Recent progress in deep learning has provided powerful new tools for inferring the intrinsic physical properties of astrophysical systems from observational data. Convolutional Neural Networks (CNNs) have been successfully applied in various astronomical contexts, particularly for identifying complex spatial structures. For example, the Convolutional Approach to Structure Identification (CASI), which is based on CNNs, has demonstrated the ability to detect stellar feedback features such as winds and outflows, and to estimate their associated mass, energy, and momentum from observations \citep{2020ApJ...905..172X,2020ApJ...890...64X}. Similarly, CNNs have been used to predict the orientation of magnetic fields by analyzing gas morphology \citep{2023ApJ...942...95X}. In addition to CNNs, other machine learning frameworks such as Denoising Diffusion Probabilistic Models (DDPMs) have also shown strong potential for recovering physical quantities from observational data that are noisy or incomplete. These models have been applied to infer a range of important parameters, including volume density \citep{2023ApJ...950..146X}, the strength of the interstellar radiation field \citep{2023ApJ...958...97X}, and magnetic field strength \citep{2025ApJ...980...52X}, often with higher accuracy than traditional inversion techniques. 

Despite the success of CNNs, their reliance on local kernels can limit their ability to capture long-range spatial dependencies in large-scale structures. This has motivated the adoption of Vision Transformers (ViTs) \citep{2020arXiv201011929D}, which utilize self-attention to prioritize global geometric context. The AstroLinformer framework recently demonstrated that transformer-augmented models outperform standard convolutional methods at recovering physical parameters from degraded images, maintaining high robustness despite significant noise and resolution loss \citep{2025Ap&SS.370...65V}.

Furthermore, as these models move toward deployment on real observational surveys, the need for rigorous Uncertainty Quantification (UQ) has become paramount to ensure scientific reliability. Modern frameworks increasingly aim to distinguish between inherent data noise (aleatoric) and model-driven uncertainty (epistemic) \citep{2017arXiv170304977K}. Probabilistic frameworks like PNet and mixture density networks are increasingly used to provide calibrated error estimates and probability density functions for astronomical parameters \citep{2023AJ....166..235S, 2023MNRAS.519.4384E}. These approaches allow researchers to navigate physical degeneracies and identify when predictions are poorly constrained by the available data.

Building on these advancements, we introduce a multi-modal deep learning framework designed to infer the physical properties of protostellar outflows from observational data. This approach leverages both spatial and spectral information, enabling the model to jointly capture the morphological and kinematic signatures of outflows. The framework is trained on synthetic observations generated from MHD simulations of massive star formation, which offer physically consistent datasets with known ground-truth parameters. By learning the mapping between observational features and intrinsic properties, the model provides a robust and scalable alternative to traditional fitting techniques.

As Paper V in this series, following Paper IV's computation of radio emission from shock-ionized structures in the outflows \citep{2024ApJ...967..145G}, this work advances automated, data-driven analysis of protostellar outflows. The paper is organized as follows: \S\ref{Data and Method} details the simulation setup, synthetic data generation, and the multi-modal machine learning architecture; \S\ref{Result} presents the results, evaluates model performance, and provides interpretable analysis of feature contributions; \S\ref{Application to Real ALMA Outflows and Uncertainty Assessment} applies the framework to real ALMA observations; and \S\ref{Conclusions} summarizes the key findings and outlines directions for future work.

\section{Data and Method}
\label{Data and Method}

\subsection{Magnetohydrodynamics Simulations}
\label{Magnetohydrodynamics Simulations}

We utilize the 3D ideal MHD simulations from \citet{2023ApJ...947...40S}, which model a disk wind outflow from a massive protostar. These simulations follow the Turbulent Core Accretion (TCA) model \citep{2003ApJ...585..850M}, with boundary conditions of the simulations tracking the self-consistent evolution of a protostar growing from 1~$M_\odot$ to over 24~$M_\odot$ within an initial 60~$M_\odot$ core embedded in a clump with $\Sigma_{\rm cl}=1\, \mathrm{g\, cm^{-2}}$ \citep{2014ApJ...788..166Z}. \xd{This bounding pressure sets an initial core radius of $R_c = 0.057$~pc. The core is threaded by a poloidal magnetic field with a total initial magnetic flux of $1~\mathrm{mG} \times R_c^2$. To manage the computational expense of resolving the inner disk over $\sim 10^5$~yr, the problem is divided into an evolutionary sequence of quasi-steady states. Because we directly analyze the simulation snapshots generated by \citet{2023ApJ...947...40S}, our evolutionary sequence intrinsically follows the physical parameters summarized in their Table 1, which details the varying accretion rates, disk radii, and wind injection properties at each protostellar mass stage.} The original simulation domain covers a single hemisphere of the outflow, extending from an injection surface located 100~au above the disk midplane to an outer boundary at a height of 26,500~au from the midplane. For the analysis in this work, we mirror the simulation domain across the disk midplane ($x_1=0$) to construct a complete, bipolar outflow. Consequently, the central $\pm 100$~au region is absent from the mirrored gas grid; however, because this missing region is entirely unresolved by the synthesized beam in our simulated ALMA observations, it does not impact the large-scale outflow morphology or the subsequent machine learning inference. A comprehensive description of the MHD simulation setup, code, and parameters can be found in \citet{2023ApJ...947...40S}. In this study, we adopt the same simulation snapshots as those used by \citet{2024ApJ...966..117X}, corresponding to protostellar masses of 2, 4, 8, 12, 16, 20, and 24~$M_\odot$.

\subsection{Synthetic Observations}
\label{Synthetic Observations}

\begin{figure*}[hbt!]
\centering
\includegraphics[width=0.98\linewidth]{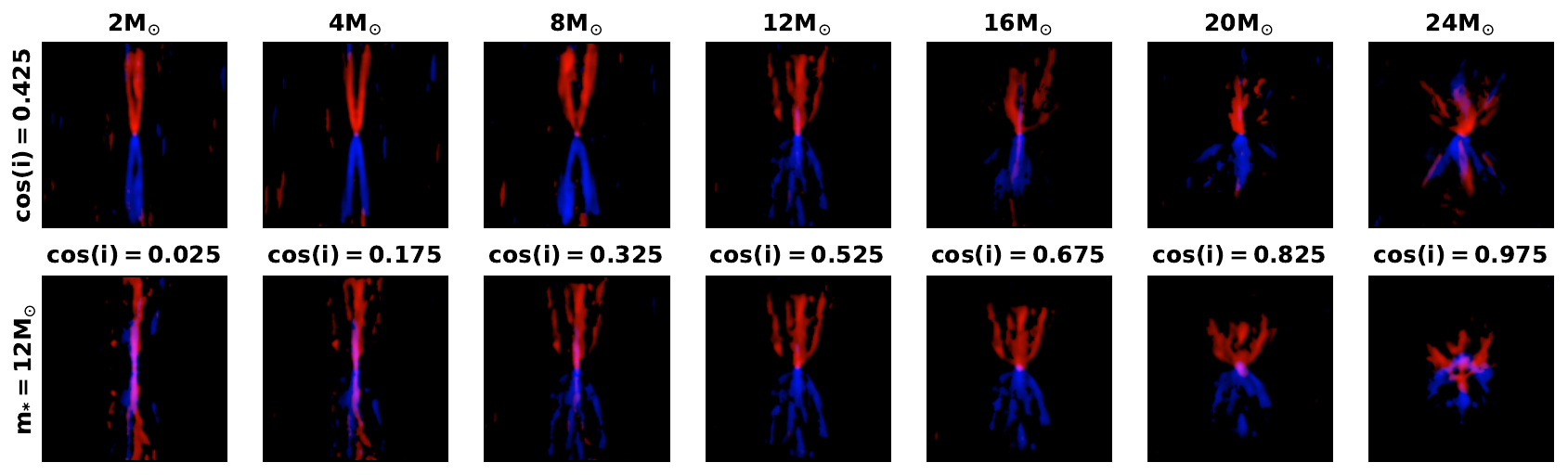}
\caption{\xd{Examples of synthetic \co\ (2-1) integrated intensity maps for outflows with different protostellar masses and inclination angles. Blue represents the blue-shifted outflow lobes ($-50<v_{blue}<-3$ \kms), while red represents the red-shifted lobes ($3<v_{red}<50$ \kms).}}
\label{fig.Outflow_Jan_ML_2chan_vel3dis2}
\end{figure*} 

\begin{figure*}[hbt!]
\centering
\includegraphics[width=0.78\linewidth]{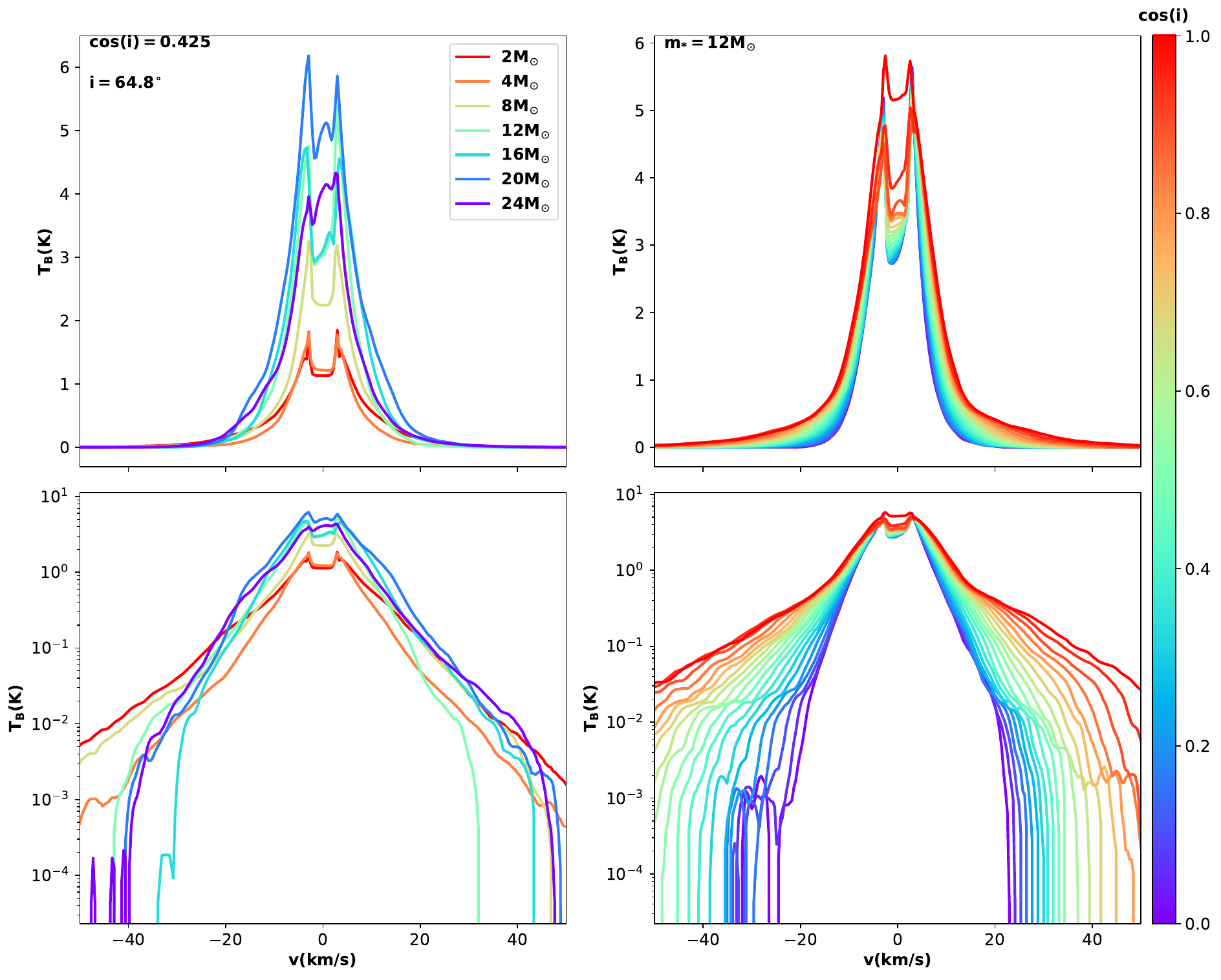}
\caption{Examples of synthetic outflow spectra for different protostellar masses and inclination angles. The upper panels show the spectra in linear scale, while the lower panels display them in logarithmic scale. }
\label{fig.Outflow_Jan_ML_spec_vel3dis2}
\end{figure*} 

We use the publicly available radiative transfer code \radmc\ \citep{2012ascl.soft02015D} to generate synthetic \co\ (2-1) spectra under non-LTE conditions. Level populations are computed based on local density and temperature, assuming that gas and dust temperatures are coupled. \xd{For these calculations, we assume a constant $^{12}$CO abundance of $10^{-4}$ relative to $H_2$ and a gas-to-dust mass ratio of 100.} This assumption is valid for the dense regions ($n_{\rm H} \gtrsim 10^5\, {\rm cm}^{-3}$) that dominate our simulation domain, and dust temperatures are calculated as described in \citet{2024ApJ...966..117X}. We also conduct radiative transfer from 20 different viewing angles, where the inclination angles are sampled by uniformly spacing the cosine of inclination angles between 0.025 and 0.975, ensuring even coverage over the spherical coordinate system. \xd{These values correspond to a range of inclination angles from approximately 12.8$^\circ$ to 88.6$^\circ$ relative to the line of sight.}

To simulate ALMA observations, we post-process the synthetic line data using CASA/\texttt{simalma}, adopting the same C36-3 array configuration and 210-second integration time as used in the real observations discussed in \S\ref{Application to Real ALMA Outflows and Uncertainty Assessment}. To diversify the training set, we include synthetic sources at distances of 500~pc, 1~kpc, and 2~kpc, using 2~kpc as the fiducial setup. While one real observed source lies at 8.4~kpc (see \S\ref{Application to Real ALMA Outflows and Uncertainty Assessment}), the model's input images and spectra are normalized, making the training process largely insensitive to distance, as also supported by \citet{2024ApJ...966..117X}. Figure~\ref{fig.Outflow_Jan_ML_2chan_vel3dis2} presents examples of synthetic outflows generated for various protostellar masses and inclination angles. Additional collections of synthetic outflow images are available in \citet{2024ApJ...966..117X}.

\xd{Similarly, Figure~\ref{fig.Outflow_Jan_ML_spec_vel3dis2} displays the corresponding synthetic $^{12}$CO (2-1) spectra for these representative models. The spectral profiles reveal that the viewing geometry plays a dominant role in shaping the observable kinematics, particularly the high-velocity line wings. As shown in the right panels for a fixed mass of $12 M_{\odot}$, varying the inclination angle ($\cos(i)$) significantly alters the spectral profile. Low-inclination (pole-on, $\cos(i) \approx 0.975$, $i \approx 12.8$\deg) configurations yield significantly broader observable velocity wings because the outflowing gas is directed predominantly along the line of sight. In contrast, high-inclination (edge-on, $\cos(i) \approx 0.025$, $i \approx 88.6$\deg ) views result in much narrower velocity profiles, as the bulk of the outflow velocity lies in the plane of the sky. While the protostellar mass also influences the overall emission intensity and line shape, reflecting changes in the underlying wind power and envelope density as the system evolves (see left panels at $\cos(i)=0.425$, $i \approx 64.8$\deg ), its impact on the maximum extent of the line wings is secondary compared to the projection effects introduced by the inclination angle. These strong, inclination-dependent spectral variations provide the multi-modal network with critical kinematic constraints that complement the spatial morphologies in the integrated intensity maps. }

The input to the machine learning model includes a three-channel image: blue-shifted and red-shifted \co\ (2-1) integrated intensities, along with the total integrated intensity, each normalized between 0 and 1. \xd{To account for missing flux near the systemic velocity, a common effect in ALMA observations, and to ensure the model is not overly sensitive to the precise definition of the central velocity, we employ a randomized velocity cutoff strategy. The blue- and red-shifted integrated intensity maps are generated using varying exclusion ranges ($|v - v_{\text{lsr}}| > 3, 5, 8,$ or $10 \text{ km/s}$), which mimics the filtering of low-velocity gas while preserving the large-scale structure of the outflow lobes. Visual inspection confirms that the resulting morphologies are robust across these ranges, allowing the network to learn consistent geometric cues despite variations in spectral coverage.} The corresponding spectrum is clipped at $\pm$50 km/s and similarly normalized. To augment the dataset, we implement several strategies in the data loader: randomly shifting the outflow center in the image, swapping the blue- and red-shifted lobes in velocity, rotating the outflow by random position angles, adding Gaussian noise to the image, and injecting pattern noise from other outflows to mimic complex structures. The added pattern noise is at least a factor of three weaker than the main signal. In addition, we introduce random spectral shifts of $\pm$5 km/s to simulate velocity-centering uncertainties commonly encountered in real observations.

\subsection{Multi-Modal Machine Learning Framework}
\label{Multi-Modal Machine Learning Framework}

\begin{figure*}[hbt!]
\centering
\includegraphics[width=0.98\linewidth]{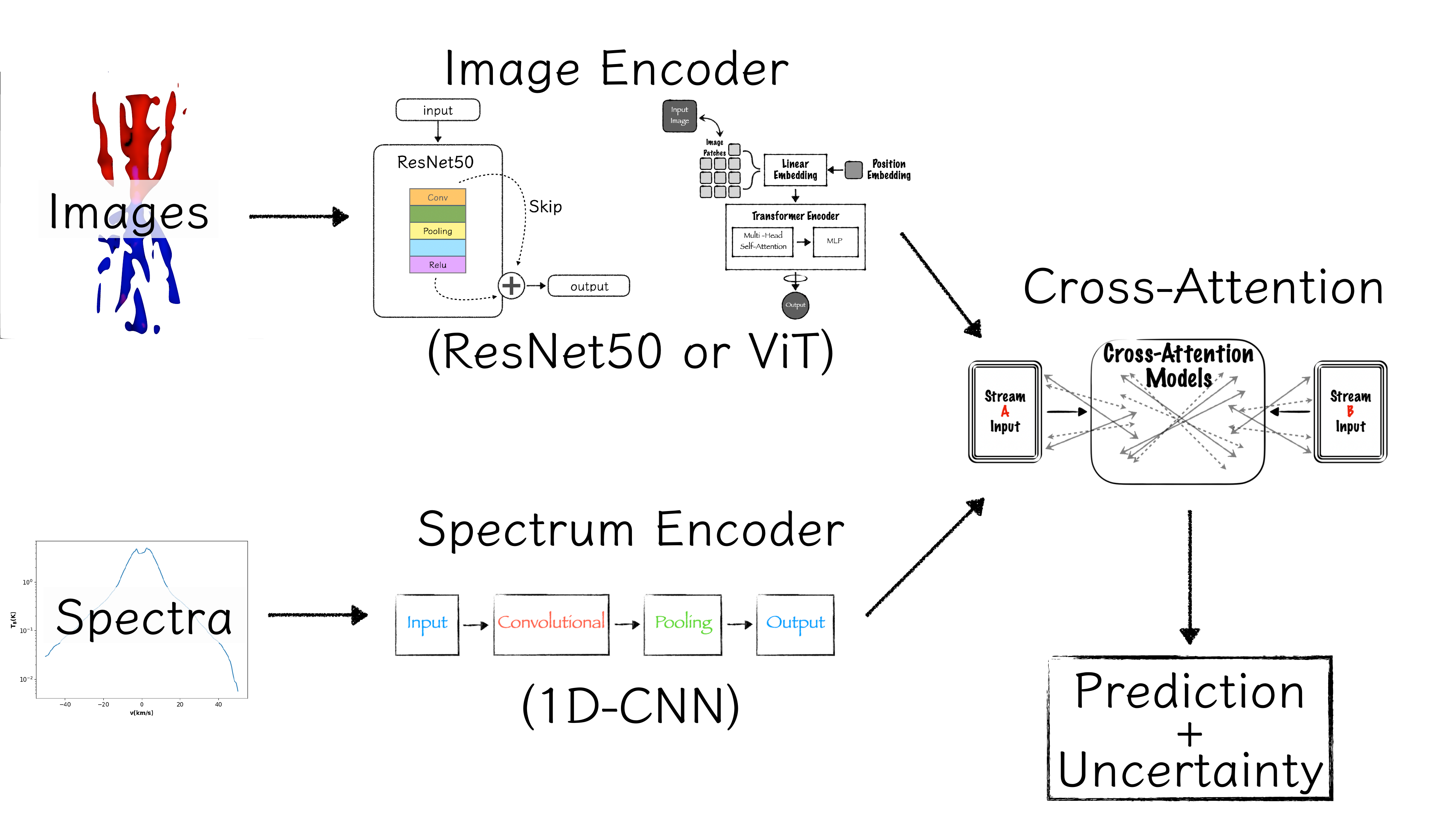}
\caption{Illustration of the multi-modal machine learning architecture. Image and spectrum encoders extract spatial and spectral features, which are fused via a cross-attention module to produce parameter predictions with associated uncertainties.}
\label{fig.ML_framework}
\end{figure*} 

We present a novel multi-modal machine learning framework that jointly processes spatial and spectral data to infer the physical properties of protostellar outflows. The framework consists of two main components: an image encoder and a spectrum encoder. For the image encoder, we adopt widely used pretrained architectures, including ResNet50 \citep{2016cvpr.confE...1H} and ViT\_L\_16 \citep{2020arXiv201011929D}, which are effective at extracting high-level spatial features from images. To enable a fair comparison across architectures, we also include ResNet152 and ViT\_B\_16, which have comparable numbers of parameters. In all cases, we use the feature representations extracted prior to the final classification layer, thereby preserving the embedded information from the input images. The spectrum encoder is designed from scratch using a 1D CNN to learn feature representations of the spectral profile, providing complementary kinematic information. To integrate these two modalities, we employ a cross-attention mechanism that fuses spatial and spectral features. The combined representation is then passed to a multi-layer perceptron (MLP), which performs the final prediction. To move beyond point estimates and quantify predictive confidence, we adopt a probabilistic regression approach. Instead of predicting a single output value, the model learns the parameters of a Gaussian probability distribution: the mean ($\mu$) and variance ($\sigma^2$). This enables the model to capture aleatoric uncertainty, the uncertainty intrinsic to the data. For example, noisy spectra or ambiguous image features naturally lead to higher predicted uncertainty (larger $\sigma$). Since the variance must always be positive, the network predicts the log-variance $s = \ln(\sigma^2)$ instead of $\sigma^2$ directly. This improves numerical stability and ensures valid outputs. The model is trained by minimizing the Gaussian Negative Log-Likelihood (NLL) of the true value $y$ given the predicted mean $\mu$ and log-variance $s$. After dropping constants, the loss function for a single prediction is:
\begin{equation}
\label{eq.NLL}
\mathcal{L}(y, \mu, s) = \frac{1}{2} \left( (y - \mu)^2 e^{-s} + s \right).
\end{equation}
This formulation encourages accurate predictions (through the $(y - \mu)^2$ term), while allowing the model to adjust its predicted variance for inputs with higher inherent uncertainty. The final model thus outputs both predictions and a principled measure of confidence. An overview of the architecture is shown in Figure~\ref{fig.ML_framework}. To further enhance robustness to real-world data imperfections, we introduce a random masking strategy in the spectrum encoder. During training, random segments of the input spectrum are set to zero, mimicking missing or corrupted spectral channels that often arise in observations.

The model is trained to predict three key physical properties for each input sample: protostellar mass, inclination angle, and position angle. The mass ranges from 2 to 24~$M_\odot$, the inclination angle from 0\deg\ to 90\deg, and the position angle from -90\deg\ to +90\deg. All outputs are normalized to improve training stability. To handle the circular nature of the position angle, we represent it as a two-component vector using sine and cosine values. This avoids discontinuities near $\pm$90\deg, which can confuse standard regression.

We follow a progressive fine-tuning approach during training. In the first stage, we freeze the image encoder and train only the spectrum encoder, cross-attention module, and MLP using a relatively high learning rate. After the validation loss fails to improve for 10 consecutive epochs, we unfreeze the last layer of the image encoder and continue training with a lower learning rate. In the final stage, we unfreeze all model parameters and reduce the learning rate further. Training continues until the validation error stops decreasing for another 10 epochs. We save the model with the best validation performance as our fiducial model.

\xdtwo{To ensure the network learns robust, physics-based representations rather than memorizing the discrete simulation grid, we implement a rigorous data splitting and dynamic augmentation strategy. To test scale invariance, our training and validation sets are strictly segregated by source distance: synthetic models placed at 1 kpc and 2 kpc are used exclusively for training, while models placed at 500 pc are reserved entirely for validation and testing. During training, we utilize an ``on-the-fly'' data loader that dynamically applies continuous augmentations to the input images and spectra at each epoch. These augmentations include adding random noise, convolving the images with random beam sizes ($\sigma$ between 0.75 and 2.0 pixels for 224$\times$224 inputs), applying random rotations ($0^\circ-360^\circ$), shifting the outflow central position to mimic real-world centering uncertainties, zooming in and out by up to 20\%, and randomly swapping the blue- and red-shifted spatial lobes while simultaneously mirroring the corresponding 1D spectrum. This dynamic pipeline ensures the model—which contains $\sim$27 million parameters in the ResNet50 configuration and $\sim$307 million in the ViT\_L\_16 configuration—is exposed to a virtually infinite number of unique sample variations, heavily mitigating the risk of overfitting. Conversely, the validation and testing sets are evaluated strictly on raw, un-augmented data. The final testing set comprises 6,720 distinct samples, representing every combination of the 7 protostellar masses, 20 inclination angles, 4 velocity cutoff ranges ($\pm 3, \pm 5, \pm 8, \text{ and } \pm 10$ km/s), and 12 fixed position angles (sampled every $30^\circ$). The training and validation loss curves, which track the Negative Log-Likelihood (NLL) and Mean Squared Error (MSE), are presented in Appendix~\ref{Training Dynamics and Learning Curves}. Our early-stopping protocol successfully captures the model weights at the global minimum before overfitting occurs; for instance, the fiducial ViT model achieved its best performance at epoch 392 with a validation NLL of -4.927 (comparable to its training NLL of -4.876) and a validation MSE of $7 \times 10^{-6}$.}


\xd{In addition to training our fiducial model on the complete physical parameter space, we construct a ``restricted dataset'' to conduct a controlled stress test of the network's generalization capabilities. In this test setup, we deliberately withhold specific physical configurations from the training phase to evaluate the model's performance on entirely unseen data. We specifically exclude all samples with a protostellar mass of $12 M_{\odot}$, as this represents an intermediate evolutionary stage, providing an ideal baseline to test whether the network can physically interpolate between lower ($8 M_{\odot}$) and higher ($16 M_{\odot}$) masses. Similarly, we exclude five evenly spaced inclination angles ($82.8^\circ, 71.0^\circ, 58.3^\circ, 43.5^\circ,$ and $22.3^\circ$, corresponding to cosine values of 0.125, 0.325, 0.525, 0.725, and 0.925) to evaluate the model's geometric interpolation across the full viewing domain. By removing these parameters, we reduce the total training sample by 35\%. This rigorous hold-out validation ensures that our assessment of the model's predictive power (discussed in Section~\ref{Assessing Model Performance}) reflects true physical interpolation rather than overfitting to the discrete simulation grid.}

\section{Results}
\label{Result}

\subsection{Assessing Model Performance}
\label{Assessing Model Performance}

\begin{figure*}[hbt!]
\centering
\includegraphics[width=0.89\linewidth]{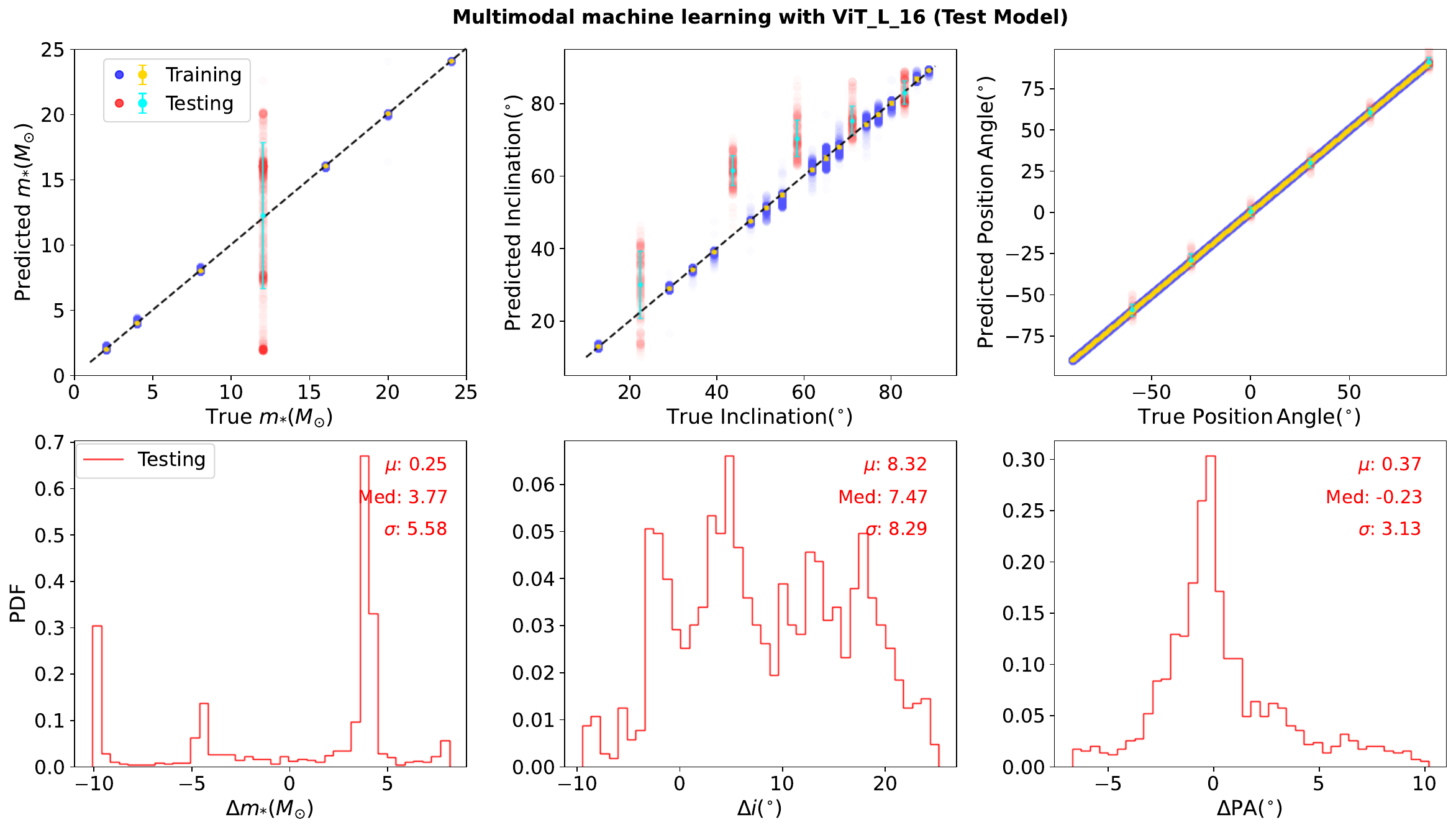}
\caption{Performance of the ViT\_L\_16 test model. See Appendix~\ref{Comparison of Performance for the Four Models} (Figure~\ref{fig.resnet_vit_12msun_test}) for a comparison of other model architectures. This model is trained on a restricted dataset that excludes samples with a protostellar mass of 12~$M_\odot$ and five specific inclination angles. Model performance is evaluated on outflows corresponding to the excluded mass and inclination angles. The top row compares the ground-truth values with the model predictions for protostellar mass, inclination angle, and position angle from left to right. The bottom row shows the probability distribution functions (PDFs) of the prediction errors, defined as the difference between the predicted and true values.}
\label{fig.vit_l_16_12msun_test}
\end{figure*} 

\begin{figure*}[hbt!]
\centering
\includegraphics[width=0.89\linewidth]{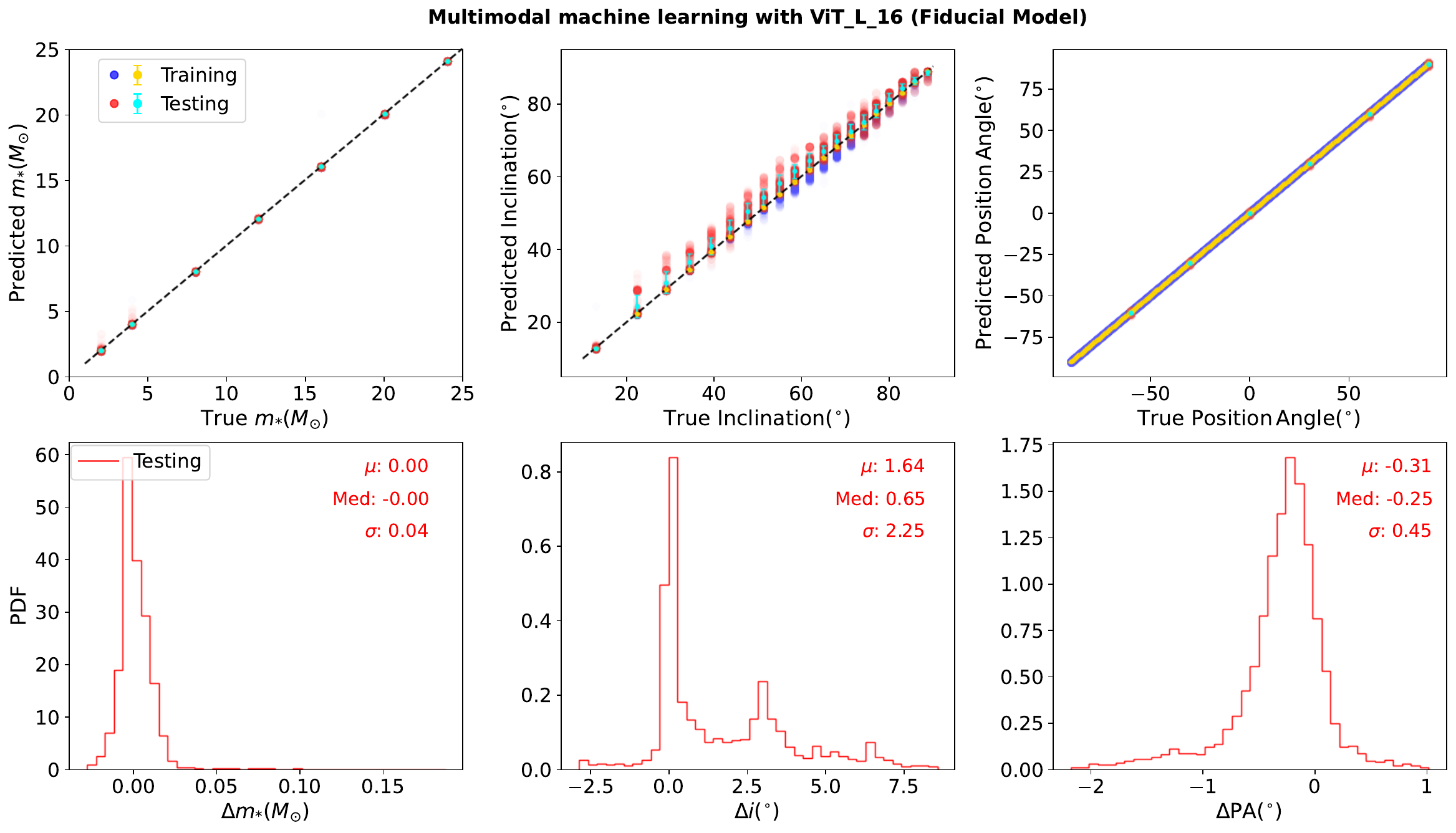}
\caption{Performance of the ViT\_L\_16 model, but trained on the full range of protostellar masses and inclination angles.}
\label{fig.vit_l_16_All_test}
\end{figure*} 

\begin{figure*}[hbt!]
\centering
\includegraphics[width=0.89\linewidth]{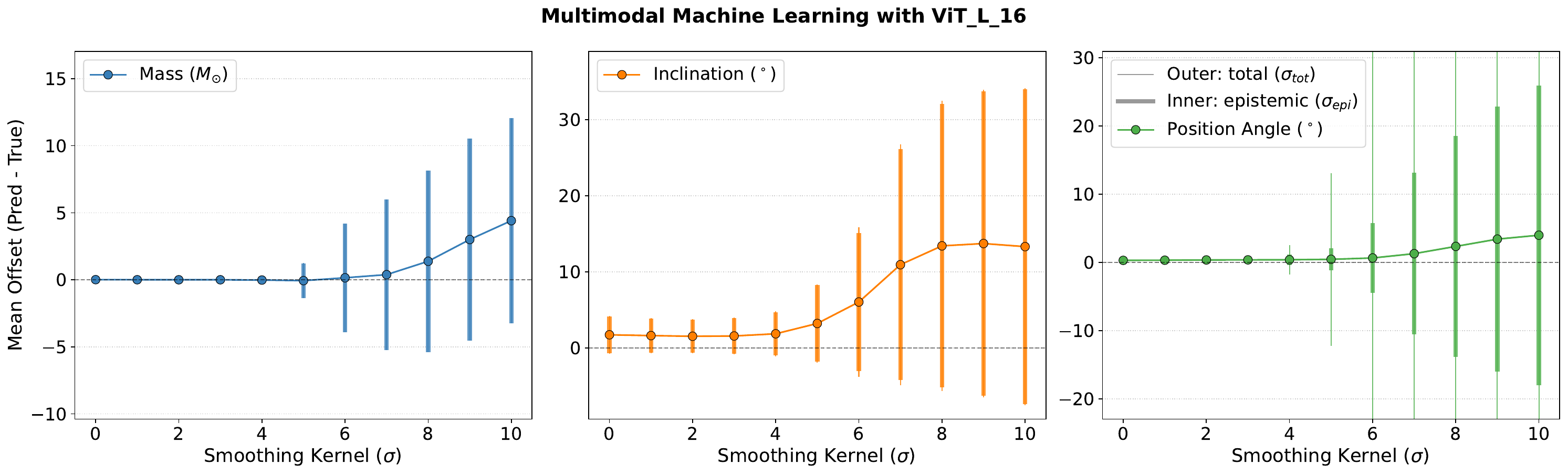}
\caption{Performance of the ViT\_L\_16 model on synthetic data convolved with progressively larger Gaussian kernels. From left to right, panels show results for protostellar mass, inclination angle, and position angle, including predicted means, total uncertainties, and epistemic (model) uncertainties. The total uncertainty is computed as the quadrature sum of epistemic and aleatoric (data) uncertainties. }
\label{fig.pred_summary_smooth_vit_l_16}
\end{figure*} 

\begin{figure*}[hbt!]
\centering
\includegraphics[width=0.89\linewidth]{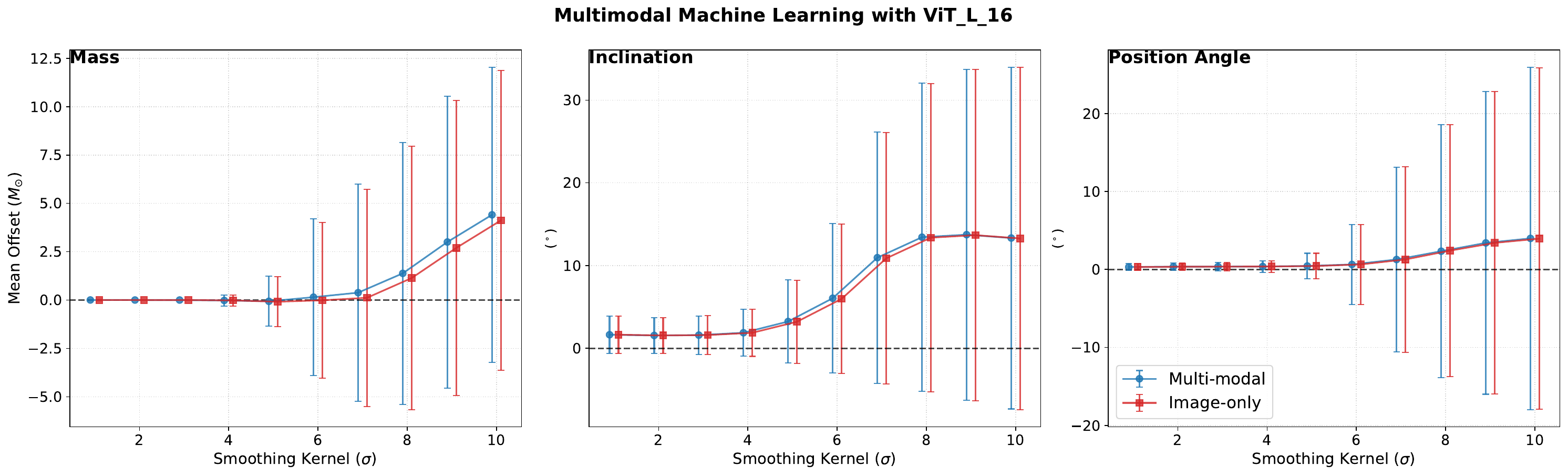}
\caption{Comparison of the ViT\_L\_16 model performance for two input configurations: multi-modal (images + spectra) and image-only, evaluated on synthetic data with progressively increased Gaussian smoothing.}
\label{fig.pred_nospectrum_test_smooth_vit_l_16}
\end{figure*}

We evaluate the performance of the trained multi-modal machine learning models using synthetic test datasets, beginning with an assessment of generalization under limited training coverage. In this setup, we exclude all samples with a protostellar mass of 12~\msun\ and five specific inclination angles from the training phase, evaluating the models exclusively on these unseen parameter combinations. \xd{This restricted setup serves as a rigorous test for generalization; by evaluating the model on physical configurations that were entirely excluded from the training set across all simulated distances, we can distinguish between the network's ability to interpolate physical trends and the mere memorization of pixel patterns from high-resolution counterparts.} Figure~\ref{fig.vit_l_16_12msun_test} illustrates the performance of the ViT\_L\_16 architecture; for conciseness, the comparative results for all four model architectures are detailed in Figure~\ref{fig.resnet_vit_12msun_test} of Appendix~\ref{Comparison of Performance for the Four Models}. A common limitation observed across all models is the discretization of predictions resulting from the sparse sampling of the physical parameter space. Consequently, predictions for the unseen 12~\msun\ case tend to cluster around the nearest training grid points (e.g., 8, 16, or 24~\msun). However, the error distributions reveal distinct behaviors between architectures: ViT-based models systematically bias predictions toward 16~\msun, suggesting an ability to capture the continuous evolutionary progression of outflow morphology. In contrast, ResNet-based models exhibit broader, less structured biases, likely reflecting intrinsic morphological degeneracies exacerbated by beam convolution and noise. \xd{We note that these structured biases are specifically a consequence of the sparse parameter coverage in this restricted test setup; as demonstrated in the following subsection, these degeneracies are effectively resolved when the architectures are trained on the full physical parameter space.} Despite these challenges in mass and inclination recovery, the position angle remains a robustly diagnosable parameter, with all four models achieving accurate recovery even for previously unseen outflow configurations.

We next evaluate the fiducial models, which are trained on the full range of protostellar masses and inclination angles using the 1~kpc and 2~kpc datasets. Testing is conducted on the reserved 500~pc dataset, utilizing a fixed set of position angles sampled at 30\deg\ intervals to ensure consistent evaluation. Figure~\ref{fig.vit_l_16_All_test} summarizes the results for the ViT\_L\_16 model; for conciseness, comparative results for all four architectures are detailed in Figure~\ref{fig.resnet_vit_All_test} of Appendix~\ref{Comparison of Performance for the Four Models}. Across the board, all models achieve accurate predictions for mass, inclination, and position angle, characterized by low bias and small uncertainties. Despite significant differences in model complexity, ranging from \xd{$\sim$27 million parameters for ResNet50 to $\sim$307 million for ViT\_L\_16}, overall performance is broadly comparable, indicating that increasing model size does not yield uniform improvements. As shown in the Appendix, ResNet152 slightly underperforms the smaller ResNet50, suggesting potential over-parameterization relative to the training data. In contrast, the ViT-based models prove more robust to scaling, with ViT\_L\_16 outperforming the ResNet architectures, particularly in protostellar mass prediction.

We then assess model robustness to resolution degradation by testing performance on synthetic data convolved with increasingly large Gaussian kernels. Figure~\ref{fig.pred_summary_smooth_vit_l_16} illustrates the performance of ViT\_L\_16, while comparative results for all four architectures are provided in Figure~\ref{fig.pred_summary_smooth_resnet_vit} of Appendix~\ref{Comparison of Performance for the Four Models}. Although training augmentations included Gaussian smoothing with $\sigma$ between 0.75 and 2.0 (for $224\times224$ inputs), we test well beyond this range. All models experience significant performance degradation when the kernel size exceeds $\sigma\sim5$, reflecting the inevitable loss of morphological information at extreme smoothing levels. However, ViT\_L\_16 consistently provides the most stable position angle predictions under heavy blurring. Similarly, ViT\_B\_16 demonstrates superior generalization compared to ResNet-based models with equivalent parameter counts. These results suggest that transformer-based architectures are particularly well-suited for handling unseen observational conditions and degraded spatial information in protostellar outflow studies.

Finally, we perform a controlled ablation test to isolate the contribution of the spectral modality. As detailed in \S~\ref{Multi-Modal Machine Learning Framework}, our training strategy involved randomly masking spectral segments, explicitly encouraging the network to prioritize spatial features. To evaluate this at inference, we disable the spectral branch entirely (setting inputs to zero) and compare performance against the standard multi-modal baseline. Figure~\ref{fig.pred_nospectrum_test_smooth_vit_l_16} illustrates the results for ViT\_L\_16, while comparative plots for all four architectures are provided in Figure~\ref{fig.pred_nospectrum_test_smooth_resnet_vit} of Appendix~\ref{Comparison of Performance for the Four Models}. The results reveal architecture-dependent sensitivities. For ResNet-based models, including spectra improves inclination angle predictions but slightly degrades mass estimates, leaving position angles largely unaffected. Conversely, ViT\_B\_16 sees a modest improvement in mass prediction from spectral data with minimal impact on other parameters. Interestingly, for ViT\_L\_16, the inclusion of spectra results in a slight performance dip for both mass and inclination. This suggests that for this larger backbone, spectral features may introduce mild confusion under the current training setup, or potentially reflects test-set variance; regardless, position angle recovery remains robust across all configurations, confirming it is driven almost exclusively by spatial morphology. Overall, under moderate blurring conditions ($\sigma \lesssim 4$), model performance remains virtually unchanged whether spectral data is included or not. This behavior is consistent with our training strategy and corroborates the Integrated Gradients analysis in \S\ref{Interpretable Analysis of Physical Feature Contributions in the Multi-Modal Network}, confirming that spatial features are the dominant drivers of the predictions.


\subsection{Interpretable Analysis of Physical Feature Contributions in the Multi-Modal Network}
\label{Interpretable Analysis of Physical Feature Contributions in the Multi-Modal Network}



\begin{figure*}[hbt!]
\centering
\includegraphics[width=0.89\linewidth]{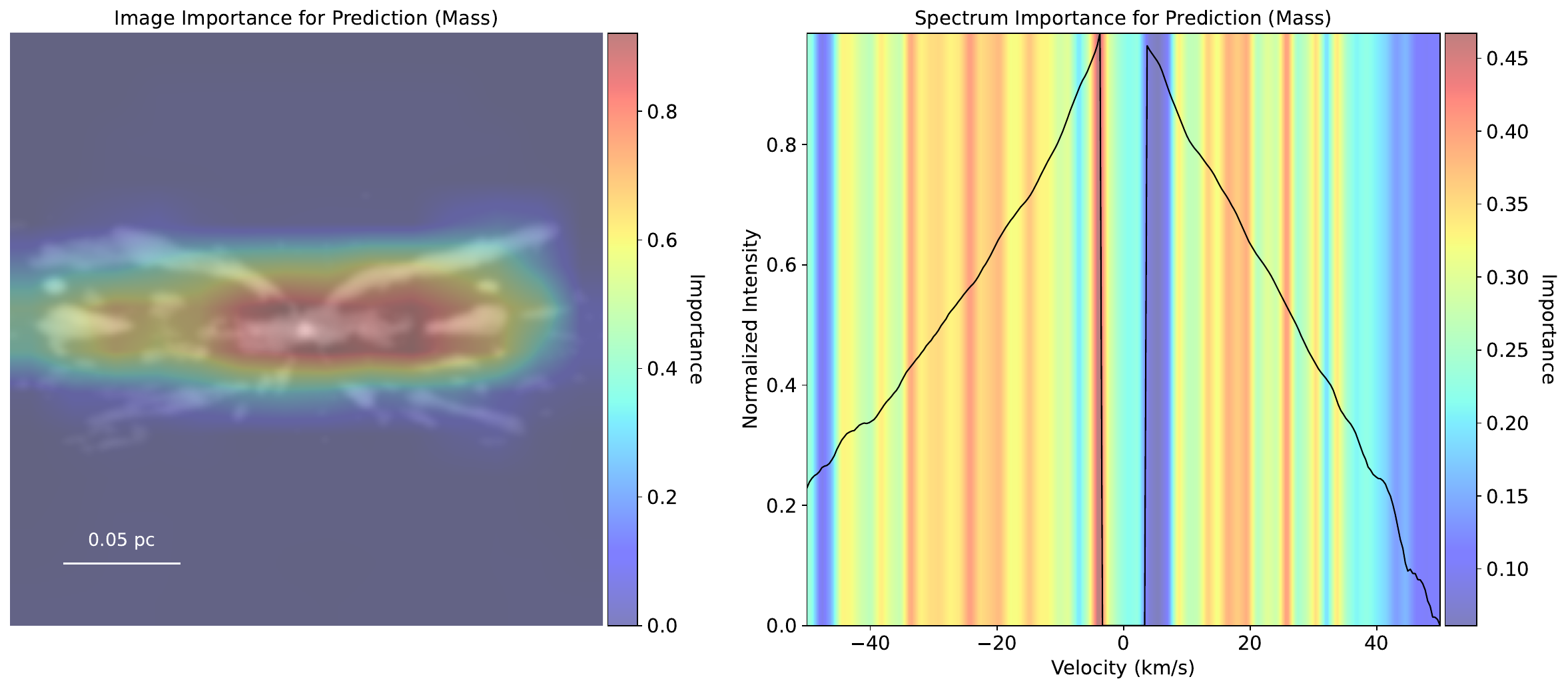}
\caption{Grad-CAM++ visualizations for outflow mass prediction using the ViT\_L\_16 model. The background shows the corresponding test outflow input, image (left) and spectrum (right). The overlaid heatmaps indicate regions that contribute most strongly to the inferred quantities. \xd{In the left panel, the spatial scale is indicated by a white scale bar representing 0.05 pc. In the right panel, the x-axis displays the line-of-sight velocity (in km s$^{-1}$), and the y-axis represents the normalized intensity (scaled from 0 to 1).} }
\label{fig.GradCAMPlusPlus_single_vit_Smooth_0}
\end{figure*}

To interpret the decision-making process of our multi-modal neural network and to quantify how specific spatial and spectral features contribute to both the predicted outflow properties (the mean, $\mu$) and their associated uncertainties (the variance, $\sigma^2$), we employ three complementary interpretability techniques: Smooth Grad-CAM++, Integrated Gradients (IG), and Occlusion Sensitivity Analysis. Together, these methods provide insight into model behavior at different levels, from coarse localization of salient regions to fine-grained attribution and direct causal testing through perturbation.

\subsubsection{Smooth Grad-CAM++ Visualization}

We first apply Smooth Grad-CAM++, an enhanced gradient-based localization method, to visualize which regions of the input images and spectra most strongly influence the model's predictions. Compared to standard Grad-CAM, the Grad-CAM++ formulation incorporates pixel-wise weighting of higher-order gradients, allowing it to better capture multiple active regions and complex spatial structures that are particularly relevant for extended objects such as protostellar outflows. To further reduce stochastic noise inherent in single-pass gradient maps, we adopt a SmoothGrad-style approach by averaging Grad-CAM++ maps over multiple noisy realizations of the input ($N=10$), with gradients explicitly enabled for our regression targets.

The resulting heatmaps highlight physically meaningful structures that drive the model's inference. These include the central outflow regions (the dense, high-velocity gas immediately surrounding the protostellar launch site), the cavity walls (the swept-up material delineating the boundary between the outflowing gas and the ambient envelope), and the terminal lobes (the extended, shocked structures at the leading edge where the outflow impacts the surrounding medium). These synthetic features correspond directly to structures commonly seen in real protostellar outflows. \xdtwo{For instance, well-defined cavity walls and terminal bow shocks are prominently observed in high-resolution studies of low-mass systems such as HH 212 \citep{2021ApJ...907L..41L, 2024ApJ...977..126L} and L1527 \citep{2017MNRAS.467L..76S}, as well as in high-mass counterparts like Orion-KL Source I \citep[e.g.,][]{2023ApJ...945...14W} and the ALMA targets analyzed in Section \ref{Application to Real ALMA Outflows and Uncertainty Assessment}.} Figure~\ref{fig.GradCAMPlusPlus_single_vit_Smooth_0} displays representative Grad-CAM++ visualizations for protostellar mass prediction using the ViT\_L\_16 model. A comprehensive comparison, including uncertainty maps and corresponding results for ResNet50, is available in Figure~\ref{fig.GradCAMPlusPlus_resnet_vit_Smooth_0} of Appendix~\ref{Gallery of Model Interpretability Visualizations}.

We find that both architectures primarily attend to the central outflow and cavity regions when predicting mass, consistent with the expectation that mass is encoded in the overall spatial extent and intensity distribution of the flow. However, distinct attention strategies emerge for position angle prediction (specifically the cosine term), as detailed in Appendix~\ref{Gallery of Model Interpretability Visualizations} (Figure~\ref{fig.GradCAMPlusPlus_resnet_vit_Smooth_2}). While the ResNet50 model focuses sharply on the edges and distal regions of the outflow cavities, ViT\_L\_16 assigns importance to a broader spatial context, including the void regions surrounding the outflow. This behavior is physically intuitive, as the outflow axis is defined as much by the geometry of the surrounding low-emission regions as by the emitting material itself.

In principle, spectral information should contribute minimally to position angle estimation, since the spectral line shape varies primarily with protostellar mass and inclination angle, while remaining invariant under rotation in the plane of the sky. However, this separation is not always clearly reflected in the Grad-CAM++ visualizations. This limitation arises from several factors. First, Grad-CAM++ operates on intermediate feature maps rather than directly on input pixels, which can obscure modality-specific contributions after multi-modal fusion. Second, the use of higher-order gradients makes the method sensitive to indirect correlations learned during training, even when those correlations are weak or non-causal. Finally, because the model jointly predicts both $\mu$ and $\sigma$, gradient signals may reflect uncertainty estimation pathways rather than direct physical dependence. These effects motivate the use of a more quantitative attribution method, which we explore next using Integrated Gradients.



\begin{figure*}[hbt!]
\centering
\includegraphics[width=0.89\linewidth]{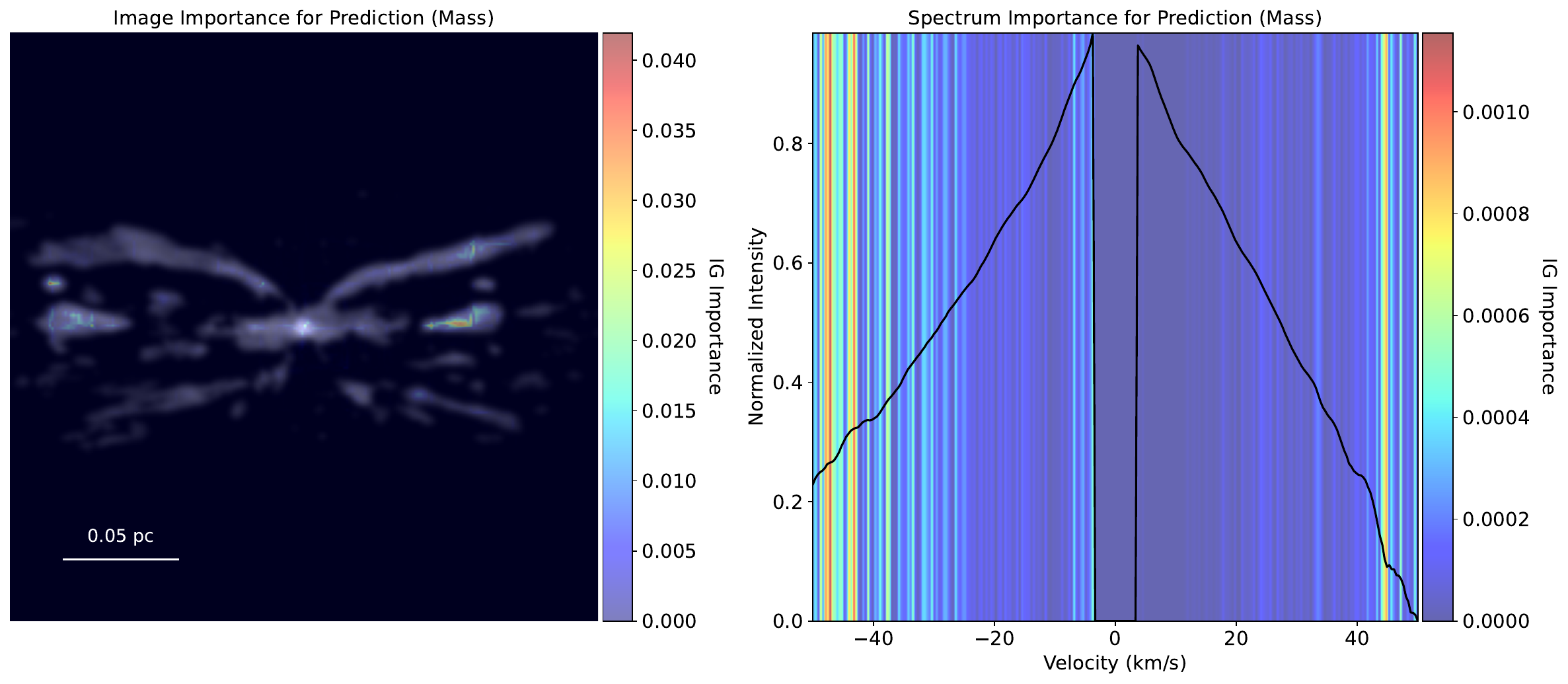}
\caption{Same as Figure~\ref{fig.GradCAMPlusPlus_single_vit_Smooth_0}, but showing Integrated Gradients maps for the prediction of outflow mass using the ViT\_L\_16 model.}
\label{fig.IG_heatmap_MANUAL_vit_single_mean_0}
\end{figure*}

\begin{figure*}[hbt!]
\centering
\includegraphics[width=0.89\linewidth]{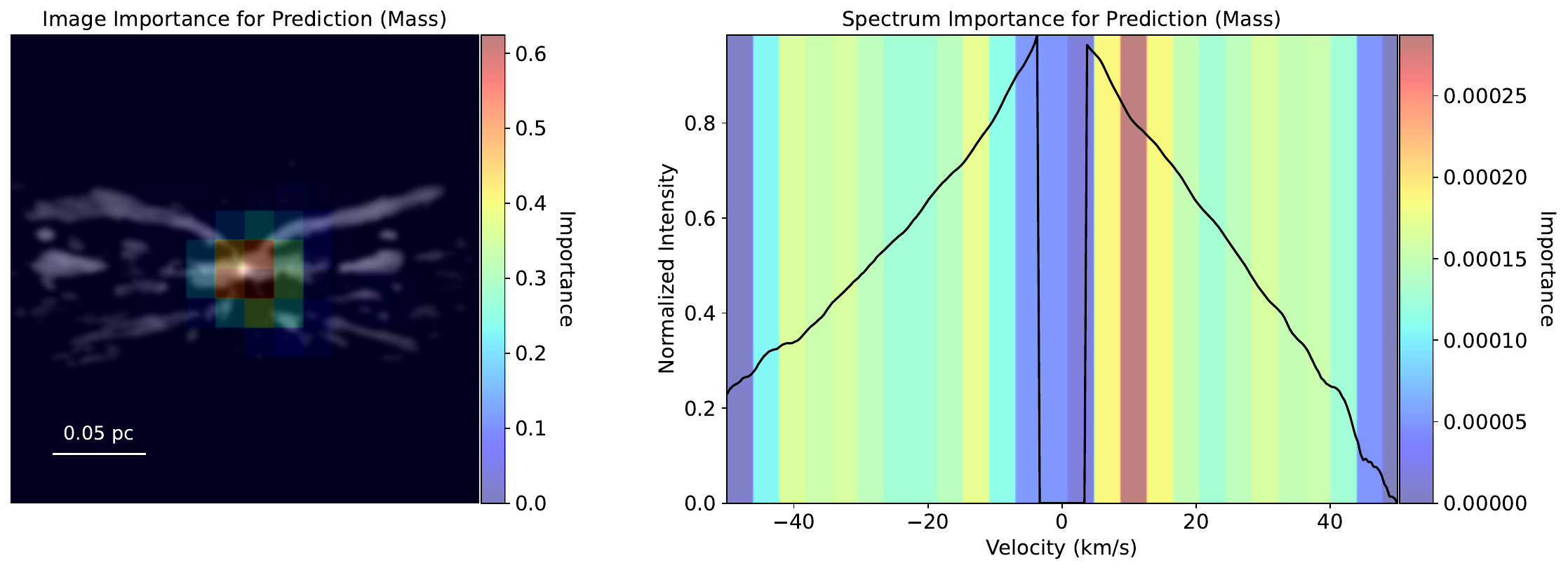}
\caption{Same as Figure~\ref{fig.GradCAMPlusPlus_single_vit_Smooth_0}, but showing Occlusion Sensitivity Analysis results for outflow mass prediction using the ViT\_L\_16 model.}
\label{fig.occlusion_uncertainty_vit_single_output_0}
\end{figure*} 

\subsubsection{Integrated Gradients with Mean Baseline}

\begin{deluxetable*}{cc|cc cc cc cc}
\tabletypesize{\footnotesize}
\tablecaption{Integrated Gradients (IG) Importance Scores by Modality and Architecture\label{tab:IG_modality_importance}}
\tablewidth{0pt}
\tablecolumns{10}
\tablehead{
\colhead{} & \colhead{} & \multicolumn{2}{c}{Mass} & \multicolumn{2}{c}{Inclination} & \multicolumn{2}{c}{PA (cos)} & \multicolumn{2}{c}{PA (sin)} \\
\colhead{Model} & \colhead{Modality} & \colhead{Mean} & \colhead{Uncert.} & \colhead{Mean} & \colhead{Uncert.} & \colhead{Mean} & \colhead{Uncert.} & \colhead{Mean} & \colhead{Uncert.}
}

\startdata
\sidehead{\textbf{ResNet50}}
& Image    & 0.0466 & 0.3176 & 0.0494 & 0.2245 & 0.0455 & 0.1861 & 0.0520 & 0.5744 \\
& Spectrum & 0.0024 & 0.1229 & 0.0092 & 0.0712 & 0.0018 & 0.0546 & 0.0027 & 0.2187 \\ \\ \hline
\sidehead{\textbf{ViT\_L\_16}}
& Image    & 0.1057 & $9.0\times10^{-6}$ & 0.0612 & 1.4397 & 0.0805 & 2.3669 & 0.0792 & 2.0430 \\
& Spectrum & 0.0011 & $1.0\times10^{-6}$ & 0.0016 & 0.4006 & 0.0001 & 0.1886 & 0.0001 & 0.2701 \\
\enddata
\tablecomments{Importance scores represent the Average $L_2$ Norm of the Integrated Gradients. The ``Uncert.'' columns refer to the importance attributed to the model's predicted $\ln(\sigma^2)$ for each physical parameter.}
\end{deluxetable*}

To obtain pixel-level and channel-level attribution with stronger theoretical guarantees, we apply Integrated Gradients (IG). This method assigns an importance score to each input feature by integrating the gradients of the output along a straight-line path from a baseline input to the actual input. Unlike gradient-only methods, IG satisfies the completeness axiom, ensuring that the sum of all attributions equals the difference between the model output and its baseline prediction.

We adopt the mean image and mean spectrum of the training dataset as the baseline, rather than a zero-valued input. This choice defines a physically meaningful reference state corresponding to an average protostellar environment, allowing attributions to be interpreted as deviations from typical conditions rather than from an artificial null input. By integrating gradients over 30-50 interpolation steps, we obtain stable and high-resolution importance maps for both spatial and spectral inputs.

Figure~\ref{fig.IG_heatmap_MANUAL_vit_single_mean_0} displays representative Integrated Gradients maps for mass prediction using the ViT\_L\_16 model. For a complete comparison, including ResNet50 results, position angle predictions, and associated uncertainty maps, refer to Figures~\ref{fig.IG_heatmap_MANUAL_resnet_vit_mean_0} and \ref{fig.IG_heatmap_MANUAL_resnet_vit_mean_2} in Appendix~\ref{Gallery of Model Interpretability Visualizations}. The analysis confirms that for both architectures, the dominant contributions stem from central outflow regions and cavity edges. While consistent with the Grad-CAM++ results, IG offers clearer modality separation, revealing that attributions associated with spectral input are systematically weaker than those derived from the images. This contrast is most pronounced in position angle predictions, where the spectral contribution is minimal, and nearly negligible for ViT\_L\_16. This behavior aligns with physical expectations, confirming that the model correctly isolates spatial morphology as the primary driver for inferring orientation on the plane of the sky.

To assess modality importance statistically, we compute the average IG attribution across the full test set of 6,720 samples for each predicted quantity. Table~\ref{tab:IG_modality_importance} summarizes the relative contributions of image and spectral inputs for the ResNet50 and ViT\_L\_16 models. In all cases, spatial information dominates the prediction of physical properties. The spectrum plays a secondary but non-negligible role in predicting protostellar mass and inclination angle, particularly for ViT\_L\_16, while its contribution to position angle prediction remains minimal for both architectures.

When examining uncertainty predictions, images again dominate the IG attributions. However, in some cases, the spectral input shows increased importance for the uncertainty associated with position angle predictions. This behavior likely reflects the structure of the probabilistic loss function, in which uncertainty estimation is coupled to residual errors in the mean prediction. When spatial cues for position angle are strong, the model may rely on subtle spectral variations to modulate its confidence, even if those variations are not directly informative for the angle itself. In addition, uncertainty pathways in the network can amplify weak correlations that are otherwise suppressed in mean predictions, leading to higher apparent spectral importance in $\sigma$ than in $\mu$.



\subsubsection{Occlusion Sensitivity Analysis}

Finally, we perform Occlusion Sensitivity Analysis to test the causal relevance of the features identified by gradient-based methods. In this approach, we systematically mask localized patches of the input image and segments of the spectrum using the dataset mean, and measure the resulting changes in both the predicted mean and uncertainty. Unlike attribution methods, occlusion directly probes how the removal of information affects model outputs.

Figure~\ref{fig.occlusion_uncertainty_vit_single_output_0} presents representative occlusion sensitivity results for mass prediction using the ViT\_L\_16 model. For a comprehensive comparison, including results for position angle and uncertainty estimates across both ResNet50 and ViT\_L\_16 architectures, refer to Figures~\ref{fig.occlusion_uncertainty_resnet_vit_output_0} and \ref{fig.occlusion_uncertainty_resnet_vit_output_2} in Appendix~\ref{Gallery of Model Interpretability Visualizations}. The analysis reveals that both architectures exhibit strong sensitivity to the masking of central outflow regions, cavity edges, and terminal lobes, confirming the causal importance of these structures for accurate inference. In contrast, occluding spectral segments produces comparatively minor changes in the model output, providing further independent support for the conclusions drawn from the Integrated Gradients analysis.

The strong agreement across Smooth Grad-CAM++, Integrated Gradients, and Occlusion Sensitivity Analysis provides robust evidence that the model's predictions are grounded in physically interpretable features of protostellar outflows. Together, these results demonstrate that the multi-modal framework not only achieves high predictive accuracy but also learns representations that are consistent with established physical intuition about outflow morphology and kinematics.

\section{Application to Real ALMA-Observed Protostellar Outflows and Uncertainty Assessment}
\label{Application to Real ALMA Outflows and Uncertainty Assessment}

In this section, we apply our trained multi-modal machine learning models to real ALMA observations of protostellar outflows in order to assess model robustness and physical plausibility beyond the synthetic domain. We analyze \co~(2-1) emission from three massive protostellar objects: G35.20−0.74N (hereafter G35.20; \citealt{2013A&A...552L..10S,2022ApJ...936...68Z}), G45.47+0.05 (hereafter G45.47; \citealt{2019ApJ...886L...4Z}), and G339.88−1.26 (hereafter G339.88; \citealt{2019ApJ...873...73Z}). Their distances are 2.2 kpc, 8.4 kpc, and 2.1 kpc, respectively.

The \co~(2-1) data were obtained with ALMA in the C36-3 configuration in 2016 (project ID: 2015.1.01454.S), with baselines ranging from 15 m to 463 m and an on-source integration time of 3.5 minutes per target. The synthesized beams have major axes of 0.87\arcsec, 0.90\arcsec, and 0.93\arcsec, and minor axes of 0.83\arcsec, 0.84\arcsec, and 0.74\arcsec\ for G35.20, G45.47, and G339.88, respectively. Data calibration and imaging were performed in CASA. After standard pipeline calibration, self-calibration using the continuum was applied to the CO line data, and imaging was carried out with the tclean task using Briggs weighting with robust = 0.5. We refer readers to the original publications for full observational details.

\begin{figure*}[hbt!]
\centering
\includegraphics[width=0.99\linewidth]{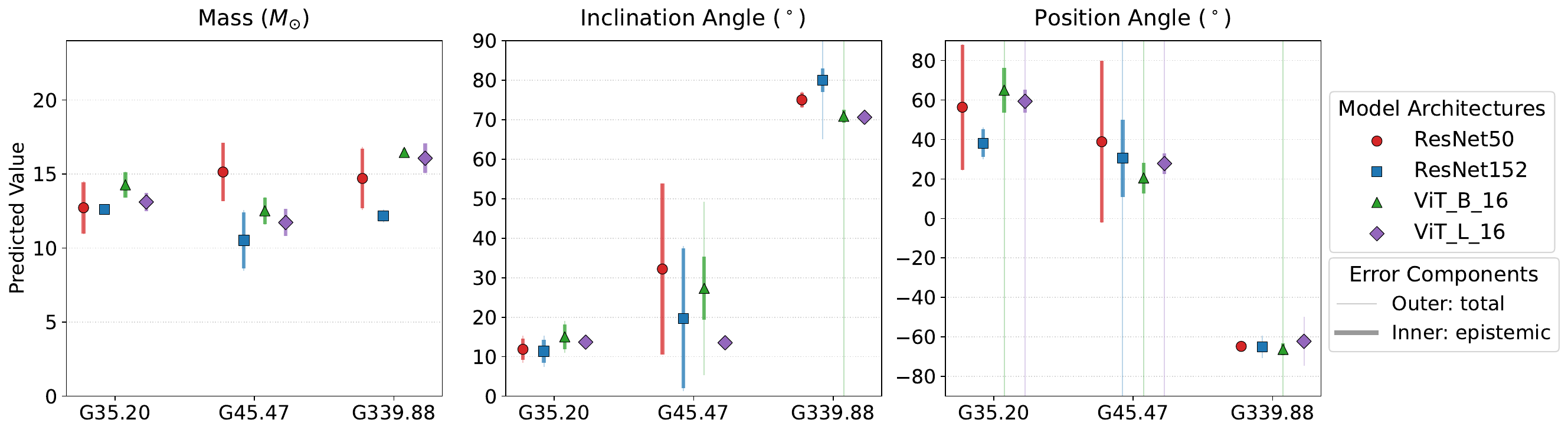}
\caption{Rotation-averaged predictions for the three ALMA-observed outflows from the four machine learning models, including the predicted means, total uncertainties, and the epistemic component arising from rotational dispersion. }
\label{fig.summary_uncertainty_outflow_alma}
\end{figure*} 

\begin{figure*}[hbt!]
\centering
\includegraphics[width=0.99\linewidth]{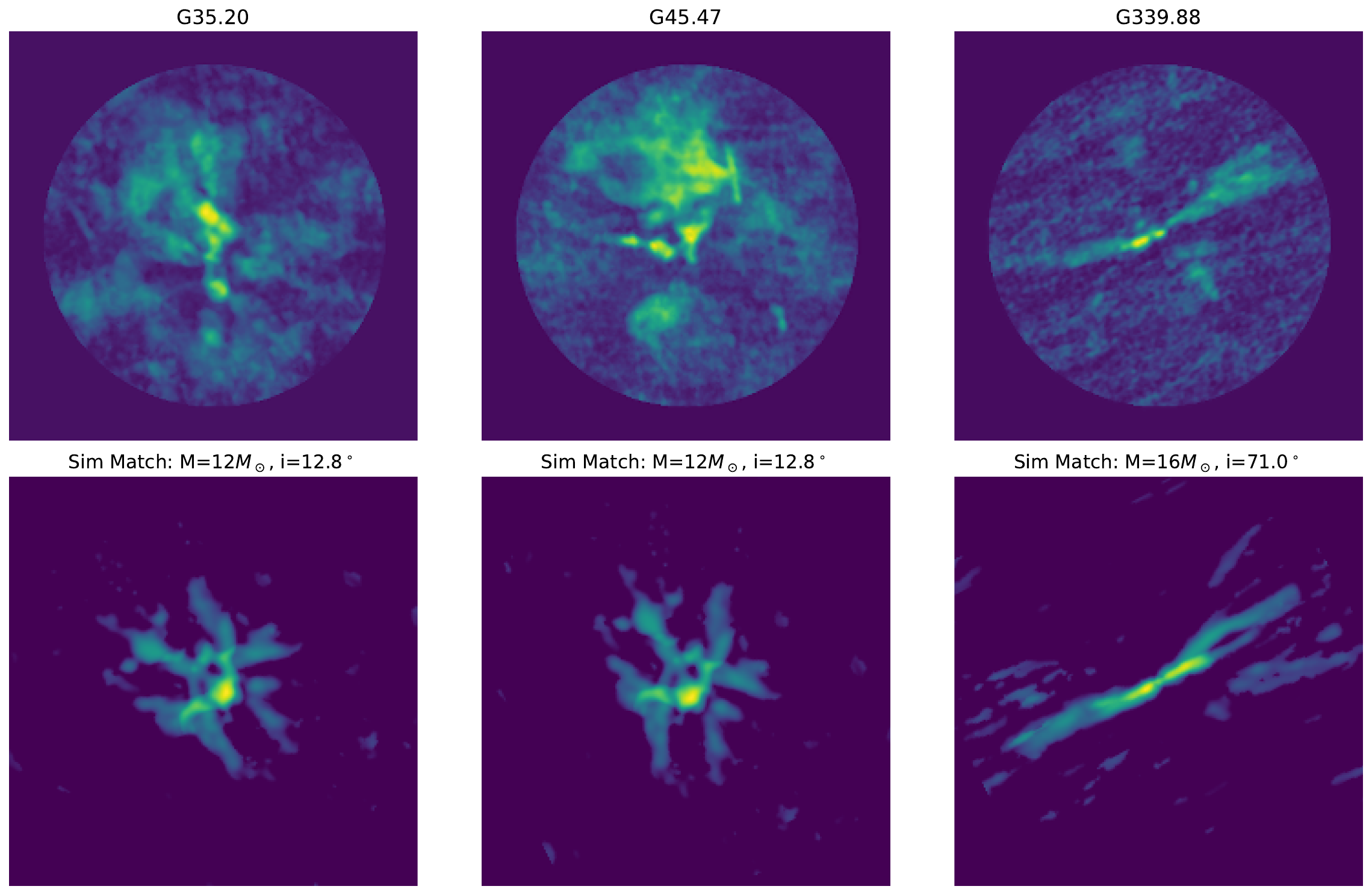}
\caption{\xd{Side-by-side comparison of real ALMA $^{12}$CO (2-1) observations (top row) and synthetic counterparts (bottom row) retrieved from the simulation grid using the model's predicted mass, inclination, and position angle. While stochastic turbulent features prevent a pixel-perfect match, the global morphology and cavity geometry show strong qualitative agreement.}}
\label{fig.obs_vs_sim_comparison}
\end{figure*}

\begin{figure*}[hbt!]
\centering
\includegraphics[width=0.99\linewidth]{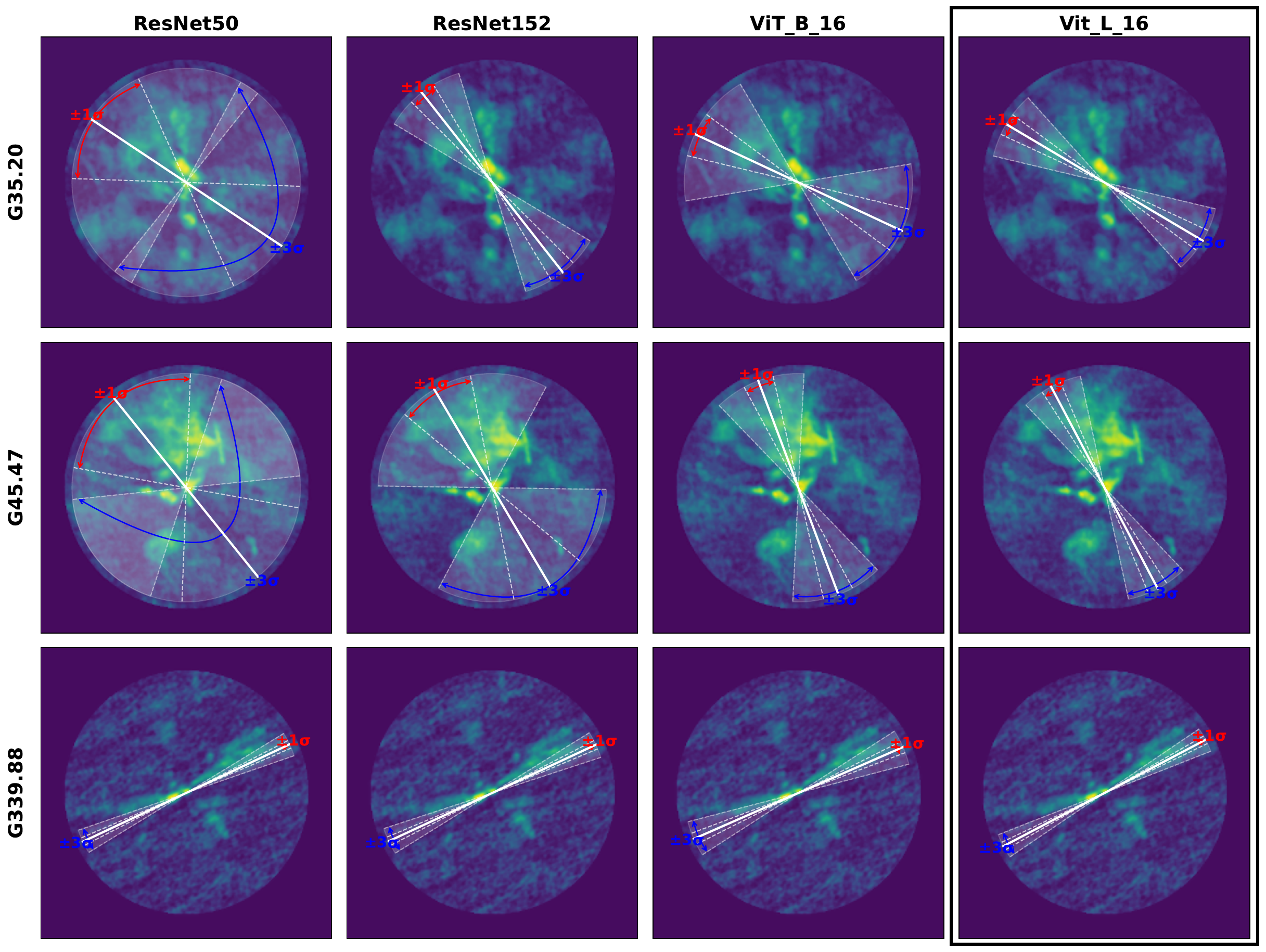}
\caption{Predicted outflow position angles for the three sources from all four machine learning models, with associated epistemic uncertainties. The background image displays the total \co~(2–1) integrated intensity, the solid white line denotes the mean predicted position angle, and the semi-transparent white wedges represent the 1$\sigma$ and 3$\sigma$ epistemic uncertainty ranges. The black box highlights the performance of the fiducial ViT model on the three sources.}
\label{fig.outflow_alma_PA_all}
\end{figure*} 

\subsection{Rotation-Based Evaluation and Uncertainty Decomposition}

To test model robustness and quantify predictive uncertainty on real observations, we exploit rotational augmentation at inference time. For each source, we rotate the outflow image by steps of 3\deg, producing 120 rotated versions that are independently passed through the trained models. This strategy enables us to separate two distinct uncertainty components:
\begin{enumerate}
\item Aleatoric uncertainty, obtained directly from the model output through the predicted $\ln(\sigma^2)$, reflects irreducible noise inherent to the data, such as finite angular resolution, instrumental noise, projection effects, and confusion from overlapping structures along the line of sight.
\item Epistemic uncertainty, estimated from the dispersion of predicted means across rotations, captures model sensitivity to input orientation and thus reflects limitations in model generalization and feature robustness.
\end{enumerate}

Ideally, physical quantities such as mass and inclination should be invariant under image rotation. Therefore, any systematic variation in predictions across rotations indicates residual orientation sensitivity in the learned representations.

Several factors can contribute to rotation-dependent prediction variability, even for a well-trained model. First, pixel interpolation during rotation introduces small but non-negligible changes in local textures and sharp edges, which can affect feature extraction. Second, standard convolution kernels and patch embeddings are not intrinsically rotation-invariant, meaning that rotated structures activate different filters or attention patterns. Third, aliasing of fine-scale ISM substructure can create artificial high-frequency features after rotation, which the network may interpret as physical morphology.

For each of the $N$ rotated inputs, the model predicts a mean $\mu_i$ and variance $\sigma_i^2$. We compute the final prediction and uncertainty as:
\begin{align}
\mu_{\rm final} &= \frac{1}{N} \sum_{i=1}^{N} \mu_i,\\
\sigma^2_{\rm total} &=
\underbrace{\frac{1}{N} \sum_{i=1}^{N} \sigma_i^2}_{\text{mean aleatoric}}
+
\underbrace{\frac{1}{N} \sum_{i=1}^{N} (\mu_i - \mu_{\rm final})^2}_{\text{epistemic (rotational dispersion)}}.
\end{align}
This formulation ensures that the total uncertainty remains large if the model is internally confident but inconsistent under rotation, providing a more realistic assessment than a single forward pass.

\subsection{Comparison with Previous Estimates}

Figure~\ref{fig.summary_uncertainty_outflow_alma} presents the rotation-averaged predictions and uncertainty decomposition for all three sources. The predicted masses for the three outflows cluster around $\sim$12-15~$M_\odot$ with relatively small total uncertainties.

For comparison, \citet{2023ApJ...942....7F} estimated protostellar masses of 13-28~$M_\odot$ for G35.20, 23-53~$M_\odot$ for G45.47, and 11-42~$M_\odot$ for G339.88 based on SED fitting, with inclination angles of approximately $60^\circ$ for all three sources. Their inferred initial core masses are significantly higher, ranging from $\sim$80 to over 400~$M_\odot$.


\xdtwo{The primary source of discrepancy is that our training simulations assume a fixed initial core mass of 60~$M_\odot$. As a result, the predicted ``protostellar mass'' ($m_*$) from our models should not be interpreted as an absolute physical mass for real systems, but rather as a proxy for the evolutionary fraction of the core ($m_* / M_{\rm core}$) within the restricted simulation parameter space. Our interpretability analysis supports this: the feature importance maps indicate that the network relies heavily on the spatial morphology of the outflow (e.g., the widening of the cavity walls) to infer the mass. However, we note that the outflow opening angle is not strictly a dimensionless geometric function of $m_* / M_{\rm core}$. Cavity widening is fundamentally a momentum-driven process that depends strongly on the wind mass-loss rate and, by extension, the absolute mass accretion rate \citep{2014ApJ...788..166Z}. Consequently, the model effectively learns the specific evolutionary sequence—and the associated momentum-injection history—of a massive 60~$M_\odot$ core . This single evolutionary track introduces an inherent degeneracy between the protostellar mass and the mass accretion rate, meaning the inference may be biased when applied to protostars embedded in significantly lower- or higher-mass envelopes with different accretion histories. Under this interpretation, the ML-inferred masses for G35.20 and G339.88 are broadly consistent with the evolutionary stages inferred from SED modeling. In contrast, G45.47 appears more evolved than suggested by SED fitting when assessed by the stellar-to-initial-core mass ratio, but less evolved in terms of absolute stellar mass, because its initial core mass inferred from SED modeling (228-444~\msun; \citealt{2023ApJ...942....7F}) is substantially larger than the fixed initial core mass adopted in our simulations.
}

The inclination predictions show more pronounced discrepancies. The models infer G35.20 and G45.47 to be relatively pole-on (inclinations $\sim$10-20\deg), whereas G339.88 is predicted to be close to edge-on (inclinations $\sim$70-80\deg). Morphologically, G339.88 indeed exhibits narrow, well-separated bipolar lobes, consistent with an edge-on configuration, whereas G35.20 and G45.47 display broader cavities and more complex structures in the training simulations associated with low-inclination outflows.

In contrast, SED-based studies tend to infer more edge-on viewing angles for G35.20 and G45.47. However, SED fitting is known to suffer from degeneracies between inclination, envelope geometry, and cavity opening angle, and may not uniquely constrain outflow orientation. Indeed, discrepancies between SED modeling and CO spectral profile fitting have also been reported, particularly for G45.47 \citep{2024ApJ...966..117X}. This highlights that neither approach should be regarded as definitive in isolation. Furthermore, in the case of G35.20, the region is known to contain multiple protostellar outflows \citep{2022ApJ...936...68Z}, which thus impact the results derived when interpreting as a single, dominant source.

\xd{To provide a direct visual validation of these results, we present Figure~\ref{fig.obs_vs_sim_comparison}, which provides a side-by-side comparison between the real ALMA $^{12}$CO (2--1) observations of our three targets and synthetic counterparts retrieved from the simulation grid using our predicted physical parameters. While some real observed outflows, such as G35.20, may only display a single prominent lobe due to environmental complexities or source multiplicity, the current machine learning model successfully identifies the underlying physical structure by finding the closest global match for mass and inclination. The qualitative agreement in lobe shape and collimation demonstrates that the model effectively maps observed morphologies back to the underlying physical properties of the disk-wind system.}

\xd{This qualitative approach to validation is necessitated by the nature of the 3D MHD simulations, in which the envelopes and outflows undergo stochastic structural evolution. Because these simulations represent specific physical realizations of a turbulent core rather than deterministic, pixel-level templates, the detailed structure of features such as cavity walls or local intensity fluctuations is not expected to match real-world sources exactly. Instead, as demonstrated in our interpretability analysis, the multi-modal framework drives its inference from robust, global morphological markers—such as cavity geometry and lobe extent—that are resolved in the integrated intensity maps. The visual consistency in these large-scale features confirms that the model is capturing the fundamental physical drivers of the outflow rather than overfitting to stochastic substructures.}

\subsection{Position Angle as a Robust Validation Metric}

A more direct and visually verifiable quantity is the outflow position angle. Figure~\ref{fig.outflow_alma_PA_all} shows the position angle predictions from all four models for the three sources, along with their epistemic uncertainties. The mean position angles are broadly consistent across models and align well with the apparent outflow orientations in the images, demonstrating that the networks have successfully learned robust morphological cues associated with bipolar geometry.

While some models exhibit large epistemic or aleatoric uncertainties for specific sources, others achieve high precision, indicating that uncertainty is strongly sample-dependent and model-dependent. The overall consistency in position angle predictions, despite domain shift from simulations to real data, provides encouraging evidence that the models capture physically meaningful outflow structures rather than overfitting to synthetic artifacts.

Together, these results suggest that while absolute physical parameters inferred from the current models should be interpreted cautiously due to limitations in the training parameter space, the models nonetheless extract robust geometric information from real observations and provide physically interpretable uncertainty estimates. This supports the feasibility of extending this framework to more realistic simulations with broader initial conditions in future work.

\xd{Beyond the physical parameters of the training set, a primary consideration for broader application is the potential domain mismatch when transitioning between observational facilities or molecular tracers. Our model is currently optimized for the specific spatial frequency and noise characteristics of ALMA interferometric data, which differ significantly from the continuous spatial sensitivity of single-dish telescopes or unfiltered simulation outputs. Furthermore, the framework is strictly calibrated for the $^{12}$CO (2-1) transition based on the specific excitation and abundance profiles of our MHD training simulations. Tracers such as SiO, SO$_2$, or HCO$^{+}$ involve distinct chemical and excitation physics, including shock-induced enhancement, selective depletion, and significantly higher critical densities, that are not captured in the current model weights. Because these species often trace higher-density gas or specific shock conditions, the mapping between their observed intensity and global parameters like mass or inclination would differ from that of the more ubiquitous CO emission. Consequently, direct application to different instruments or tracers could result in performance degradation due to these differing physical and instrumental responses. To ensure high fidelity across diverse setups, a ``fine-tuning'' or transfer learning step would be required. This involves re-calibrating the pre-trained model using a specialized synthetic dataset processed through the target instrument’s specific response function or the appropriate radiative transfer physics of a new molecular tracer, allowing the network to adapt to new conditions while preserving the fundamental physical correlations learned from the underlying simulations.} 

\section{Conclusions}
\label{Conclusions}

We have developed a multi-modal machine learning framework that jointly exploits spatial and spectral information to infer the physical properties of protostellar outflows. By coupling image and spectrum encoders through a cross-attention fusion mechanism, the model integrates morphological and kinematic features in a physically motivated manner. The framework was trained and validated on synthetic observations and successfully applied to real ALMA $^{12}$CO (2-1) data. Our main findings are summarized as follows:

\begin{enumerate}


\item \xd{Across all four backbone architectures evaluated (ResNet50, ResNet152, ViT\_B\_16, and ViT\_L\_16), the framework accurately recovers protostellar properties; specifically, when trained on the full parameter space, all models exhibit low systematic bias, whereas tests on restricted datasets reveal that Transformer-based architectures are more robust to morphological degeneracies than ResNet models.}

\item  Model performance degrades gradually as spatial resolution is reduced via Gaussian smoothing. However, Vision Transformer-based models, particularly ViT\_L\_16, demonstrate superior robustness under strong blurring. This suggests that the self-attention mechanism is more effective than traditional convolutions at preserving global outflow geometry when fine-scale morphological structures are suppressed.

\item Interpretability analyses using Smooth Grad-CAM++, Integrated Gradients, and Occlusion Sensitivity reveal a physically consistent dependency: spatial features dominate the prediction of all parameters, while spectral profiles provide critical secondary constraints for mass and inclination. The minimal contribution of spectra to position angle estimation aligns with physical expectations, confirming that the model has learned meaningful modality-specific physical correlations.

\item When trained on sparse parameter spaces, ViT-based models show a significantly stronger ability to interpolate physical trends compared to ResNet architectures. This suggests that Transformers are better suited for capturing the continuous evolution of outflows rather than merely memorizing discrete sampled values.

\item By employing a rotation-based uncertainty decomposition on ALMA data, we successfully separated aleatoric (data-driven) and epistemic (model-driven) contributions. We find that mass estimates remain relatively stable, whereas inclination and position angle exhibit larger epistemic scatter in complex sources, providing a more rigorous and realistic measure of model confidence than single-pass predictions.

\item For the analyzed ALMA sources, inferred protostellar masses cluster between $12\text{-}15~M_{\odot}$. These values are lower than traditional SED-based estimates, reflecting the $60~M_{\odot}$ initial core mass constraint of the training set; thus, these results are best interpreted as indicators of the evolutionary stage. While recovered position angles show excellent agreement with visual morphology, inclination angles exhibit larger discrepancies, highlighting the inherent degeneracies between viewing geometry and evolutionary effects.

\end{enumerate}

\xdtwo{Overall, this work demonstrates that multi-modal learning provides a physically interpretable and scalable approach for linking observed outflow signatures to underlying protostellar properties. However, it is vital to explicitly acknowledge the physical limitations inherent to the current training dataset. Most notably, because the underlying MHD simulations utilize a fixed initial core mass of $60~M_\odot$, the model is trained on a single evolutionary track. Consequently, the predicted protostellar mass ($m_*$) is degenerate with the mass accretion rate and primarily acts as a proxy for the evolutionary stage ($m_* / M_{\rm core}$) of a massive core, rather than representing a universal absolute mass. Furthermore, the current framework is trained exclusively on synthetic $^{12}$CO (2-1) emission. While bounded by these physical and observational constraints, the framework is modular. Future expansions of the simulation grid to include varied initial core masses and envelope densities, combined with targeted transfer learning for additional molecular tracers, will break these physical degeneracies. Ultimately, these developments will facilitate the automated, physically informed analysis of large-scale ALMA surveys and future high-resolution facilities.}

\begin{acknowledgments}

This paper is dedicated to Jan E. Staff (29th May 1977 - 16th May 2023). \xd{We thank the anonymous referee for their careful reading of the manuscript and for their constructive and insightful comments. Their suggestions have significantly improved the clarity and robustness of this work.} D.X. acknowledges support from the Virginia Initiative on Cosmic Origins (VICO) and the Natural Sciences and Engineering Research Council of Canada (NSERC), [funding reference number 568580]. D.X. also acknowledges support from the Eric and Wendy Schmidt AI in Science Postdoctoral Fellowship Program, a program of Schmidt Sciences. I.A.S. acknowledges support from a Chalmers Astrophysics and Space Sciences Summer (CASSUM) research fellowship. J.C.T. acknowledges support from NSF grant AST-2206450 and ERC Advanced grant 788829 (MSTAR). J.C.T also acknowledges funding from the Virginia Institute for Theoretical Astrophysics (VITA), supported by the College and Graduate School of Arts and Sciences at the University of Virginia. Computations were performed on the Trillium supercomputer at the SciNet HPC Consortium. SciNet is funded by Innovation, Science and Economic Development Canada; the Digital Research Alliance of Canada; the Ontario Research Fund: Research Excellence; and the University of Toronto
\end{acknowledgments}


\bibliographystyle{aasjournalv7}
\bibliography{references}

\appendix

\section{Training Dynamics and Learning Curves}
\label{Training Dynamics and Learning Curves}

\begin{figure*}[hbt!]
\centering
\includegraphics[width=0.5\linewidth]{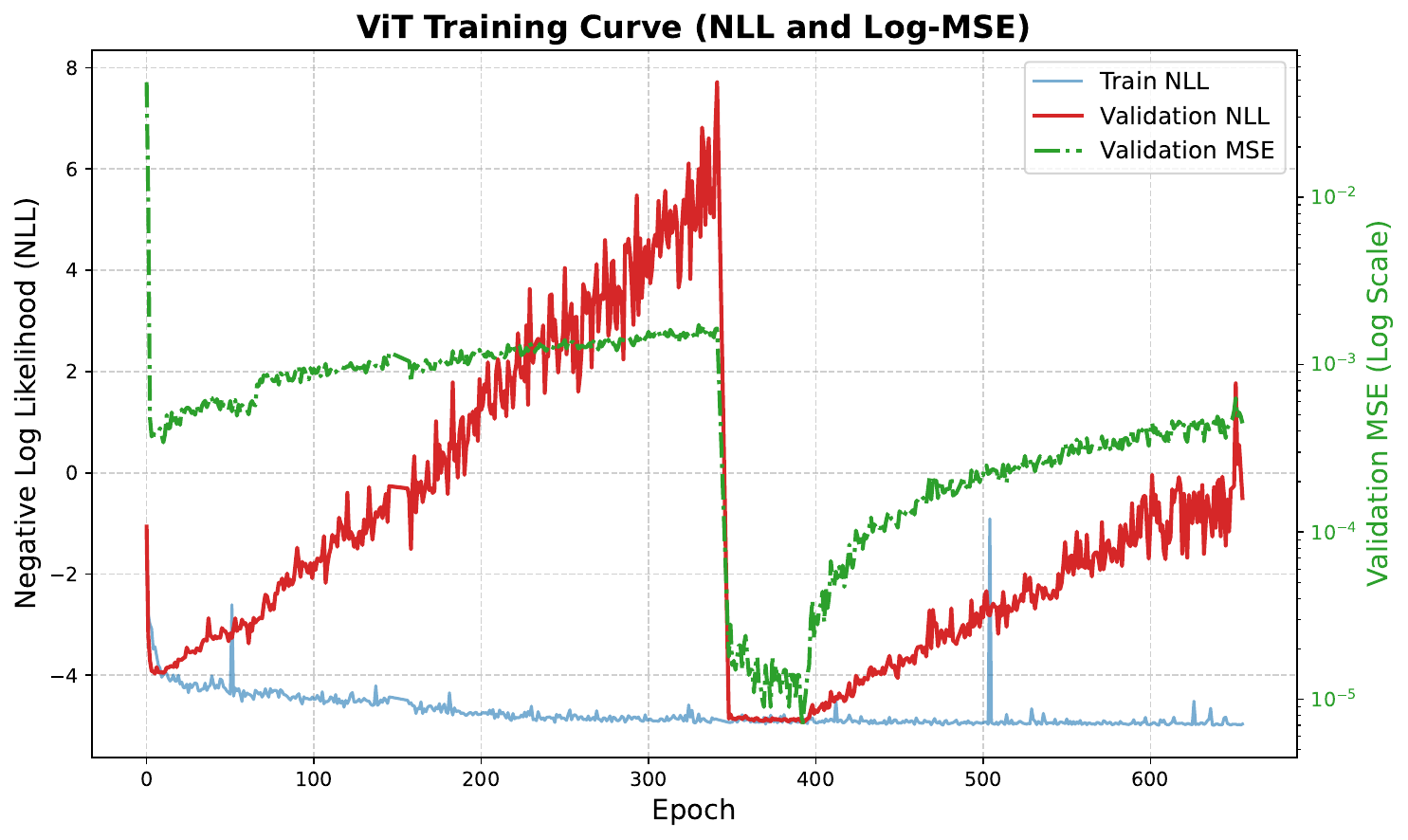}
\caption{Learning curve for the ViT\_L\_16 multi-modal model. The plot shows the training NLL (blue solid line), validation NLL (red solid line), and validation MSE (green dashed line, logarithmic scale on the right axis) as a function of training epoch. The sharp drop near epoch 350 corresponds to a scheduled learning rate decay and layer unfreezing, leading to the optimal model weights saved at Epoch 392.}
\label{fig.vit_learning_curve}
\end{figure*}

\begin{figure*}[hbt!]
\centering
\includegraphics[width=0.5\linewidth]{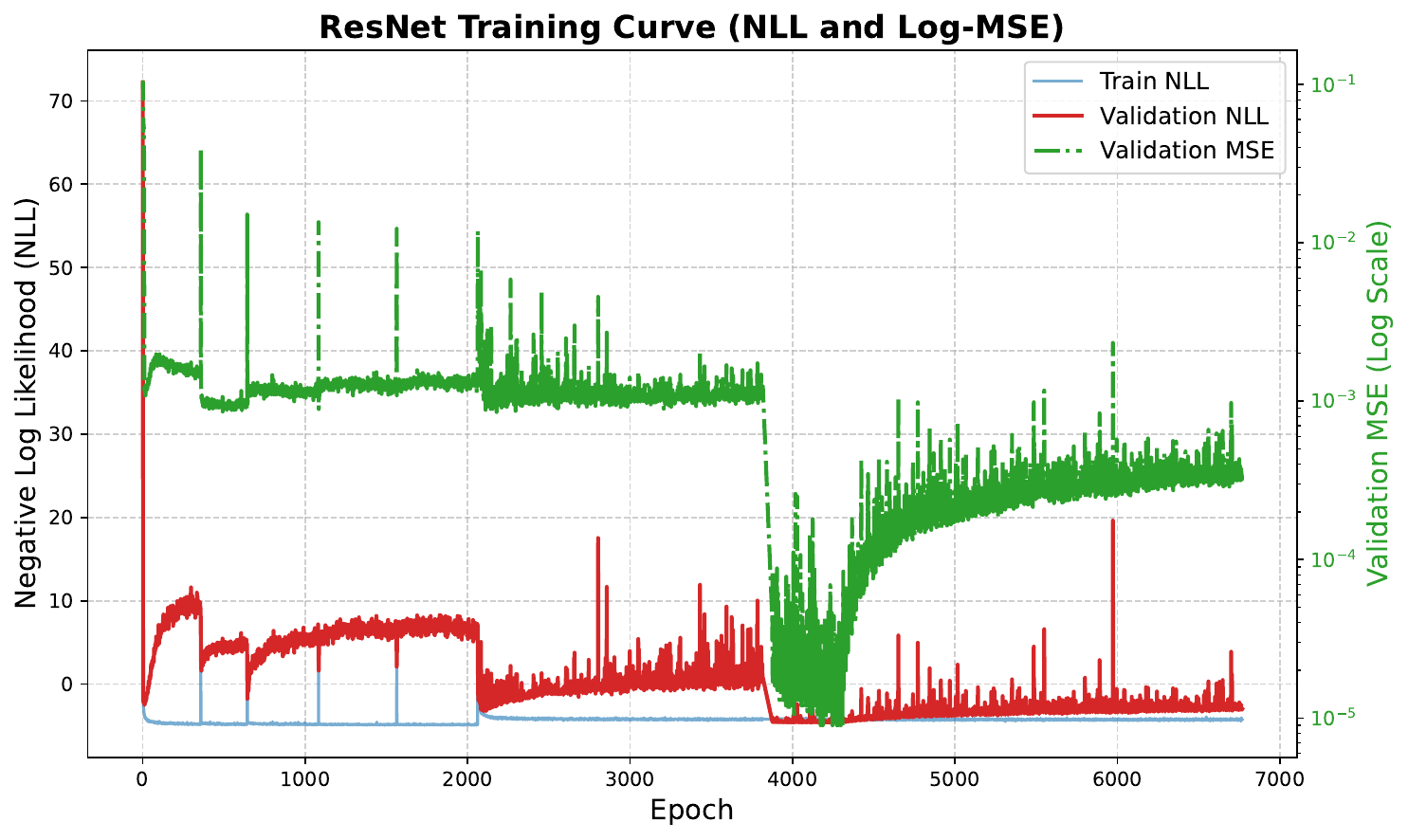}
\caption{Learning curve for the ResNet50 multi-modal model. The plot shows the training NLL (blue solid line), validation NLL (red solid line), and validation MSE (green dashed line, logarithmic scale on the right axis). The model successfully recalibrates after progressive fine-tuning milestones, achieving its lowest validation loss at Epoch 4273.}
\label{fig.resnet_learning_curve}
\end{figure*}

\xdtwo{To provide transparency regarding the model training process and to demonstrate the convergence stability of our multi-modal framework, we present the learning curves for both the ViT\_L\_16 and ResNet50 architectures. Figures~\ref{fig.vit_learning_curve} and \ref{fig.resnet_learning_curve} display the evolution of the Gaussian Negative Log-Likelihood (NLL) for both the training and validation sets, alongside the validation Mean Squared Error (MSE) plotted on a logarithmic scale. }

\xdtwo{As detailed in Section \ref{Multi-Modal Machine Learning Framework}, our training procedure utilizes a progressive fine-tuning strategy combined with heavy, on-the-fly data augmentation. During the initial training phases, the models rapidly optimize the NLL loss on the training data. Because the NLL loss function evaluates both the predicted mean and the predicted log-variance ($s = \ln(\sigma^2)$), the network can initially become overly confident, predicting artificially narrow variances for the training set. When applied to the unseen validation set, this overconfidence temporarily drives up the validation NLL, despite the underlying validation MSE remaining relatively stable.}

\xdtwo{However, at the scheduled learning rate decay and layer unfreezing milestones, the network successfully recalibrates its uncertainty estimations. This recalibration is visually evident in the learning curves as sudden, sharp drops in the validation NLL, accompanied by corresponding decreases in the validation MSE. These drops indicate that the models have escaped local minima and successfully generalized to the validation data.}

\xdtwo{Our early-stopping protocol is designed to capture the model weights exactly at these global minima, halting the training process before long-term overfitting can degrade performance. For the fiducial ViT\_L\_16 framework (Figure~\ref{fig.vit_learning_curve}), the lowest validation NLL was achieved at Epoch 392, yielding a validation NLL of $-4.93$, an associated training NLL of $-4.88$, and a validation MSE of $7 \times 10^{-6}$. Similarly, the ResNet50 framework (Figure~\ref{fig.resnet_learning_curve}) reached its optimal performance at Epoch 4273, with a validation NLL of $-4.61$, a training NLL of $-4.25$, and a validation MSE of $9 \times 10^{-6}$. The tight alignment between the training and validation NLL at these specific stopping epochs confirms that the networks are well-calibrated and have learned robust physical representations rather than memorizing the discrete simulation grid.}

\section{Comparison of Performance for the Four Models}
\label{Comparison of Performance for the Four Models}

\begin{figure*}[hbt!]
\centering
\includegraphics[width=0.49\linewidth]{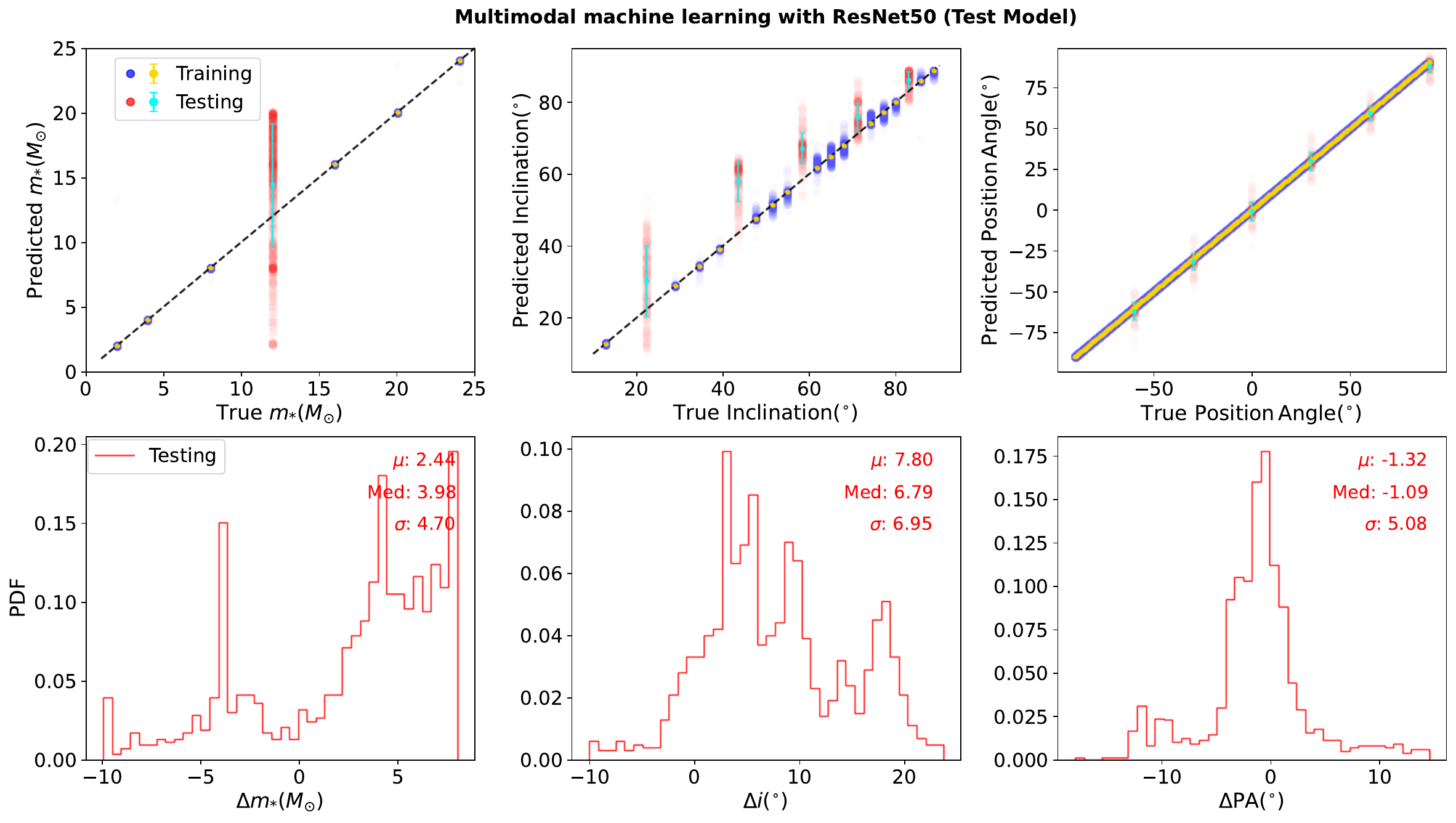}
\includegraphics[width=0.49\linewidth]{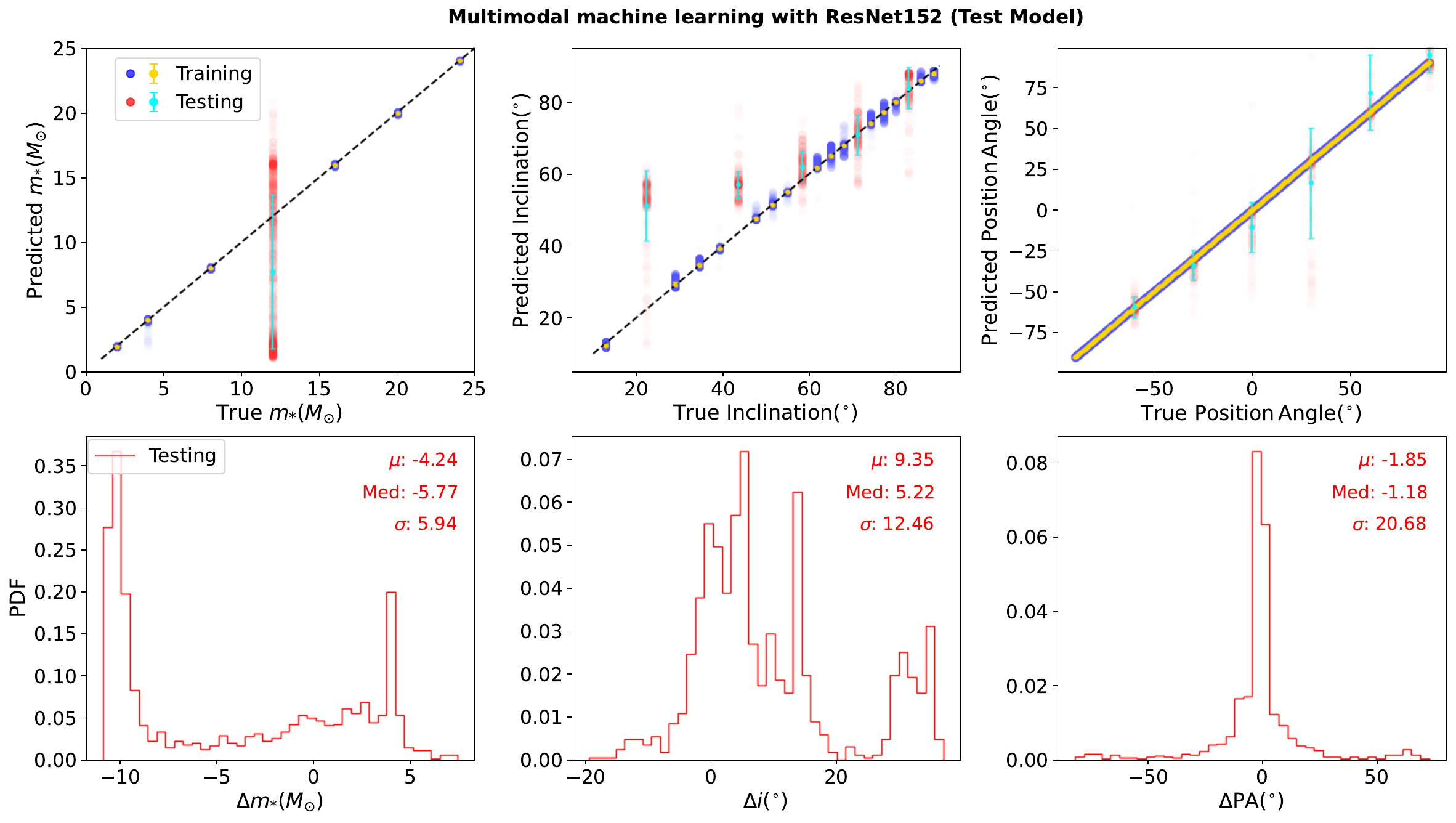}
\includegraphics[width=0.49\linewidth]{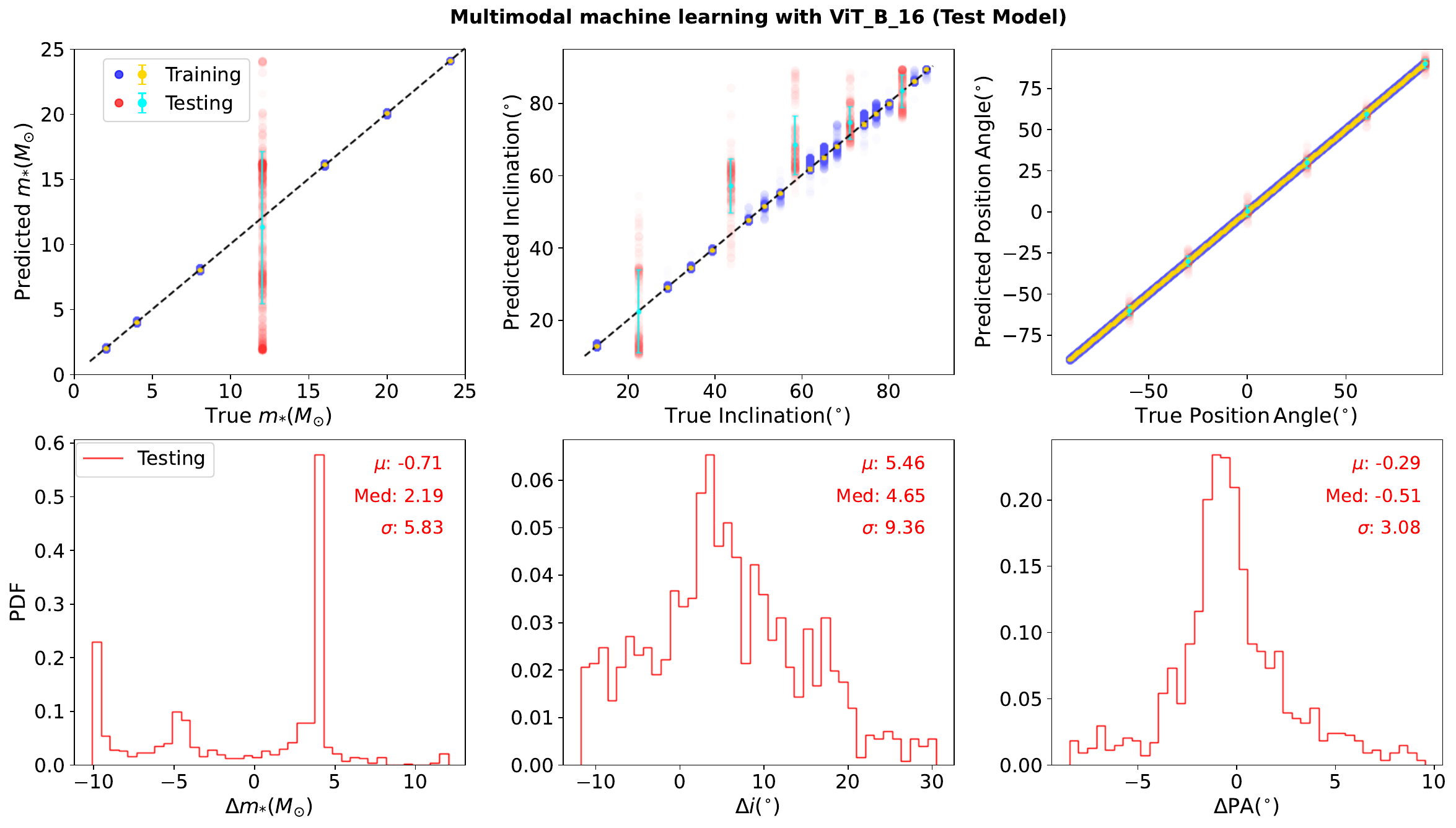}
\includegraphics[width=0.49\linewidth]{{pred_vit_l_16_uncertainty_test_12msun_0628}.pdf}
\caption{Summary of the performance of four test machine learning models: ResNet50 (upper left), ResNet152 (upper right), ViT\_B\_16 (lower left), and ViT\_L\_16 (lower right). These models are trained on a restricted dataset that excludes samples with a protostellar mass of 12~$M_\odot$ and five specific inclination angles. Model performance is evaluated on outflows corresponding to the excluded mass and inclination angles. In each panel, the top row compares the ground-truth values with the model predictions for protostellar mass, inclination angle, and position angle from left to right. The bottom row shows the probability distribution functions (PDFs) of the prediction errors, defined as the difference between the predicted and true values.}
\label{fig.resnet_vit_12msun_test}
\end{figure*} 

\begin{figure*}[hbt!]
\centering
\includegraphics[width=0.49\linewidth]{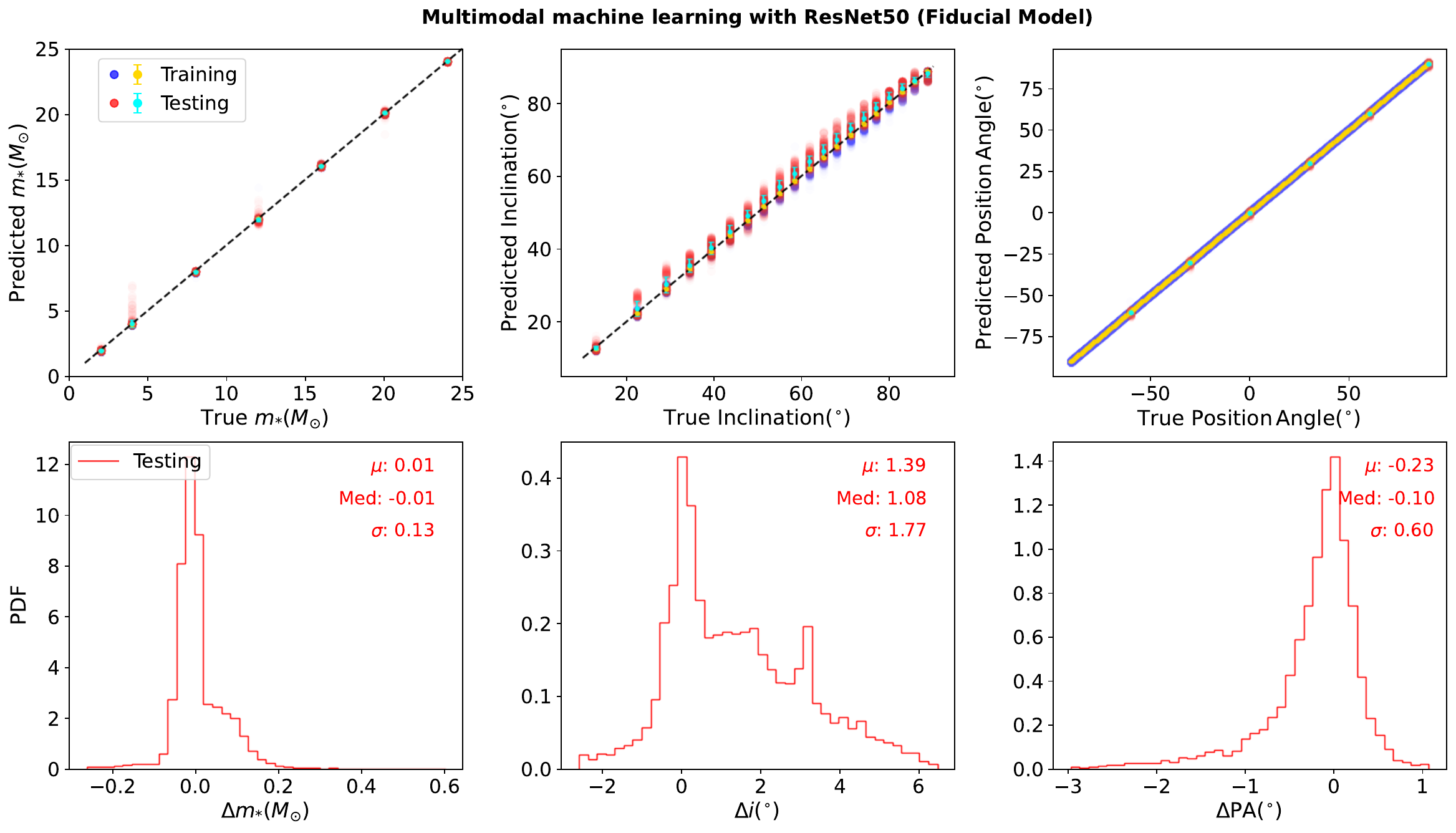}
\includegraphics[width=0.49\linewidth]{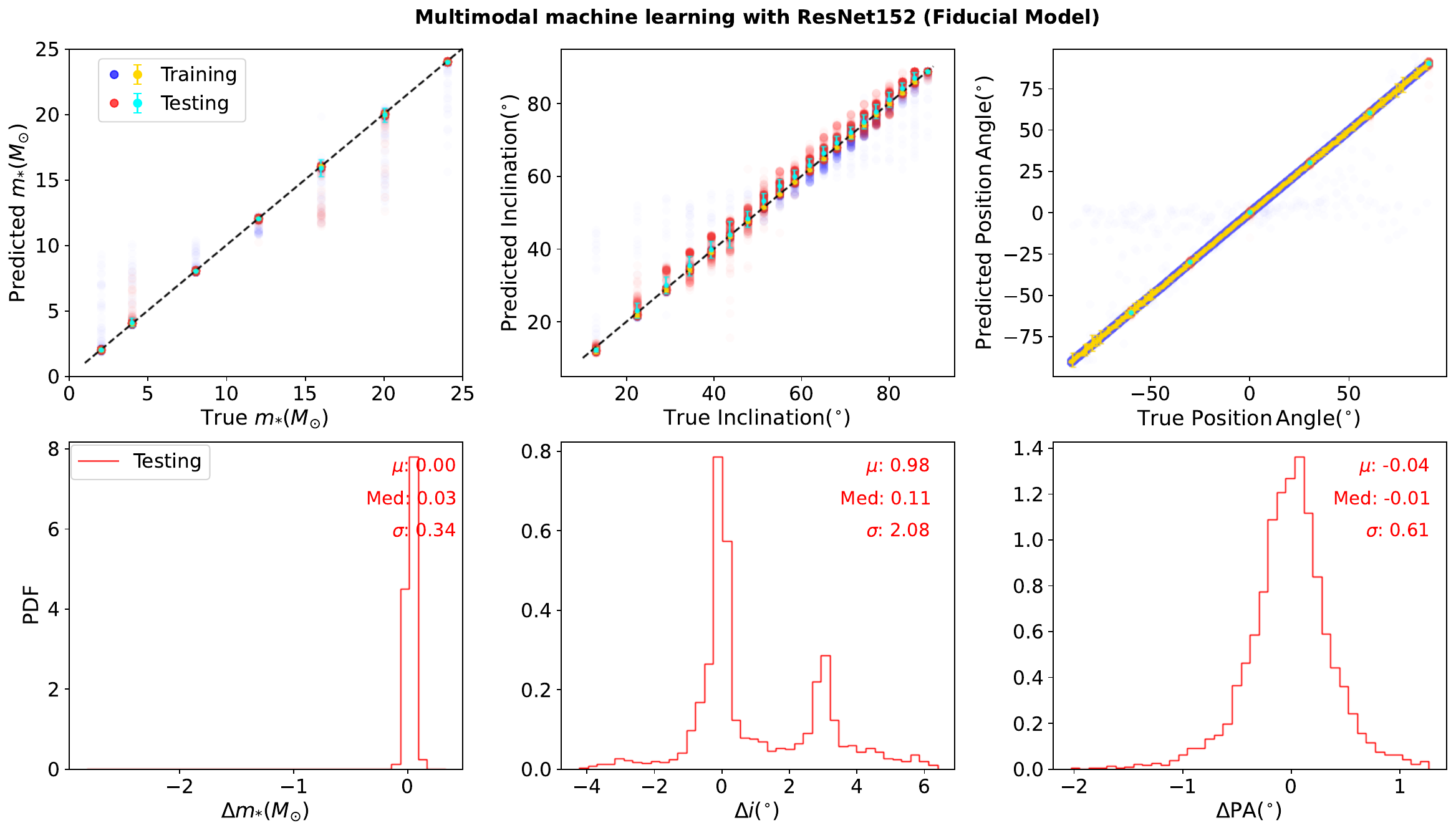}
\includegraphics[width=0.49\linewidth]{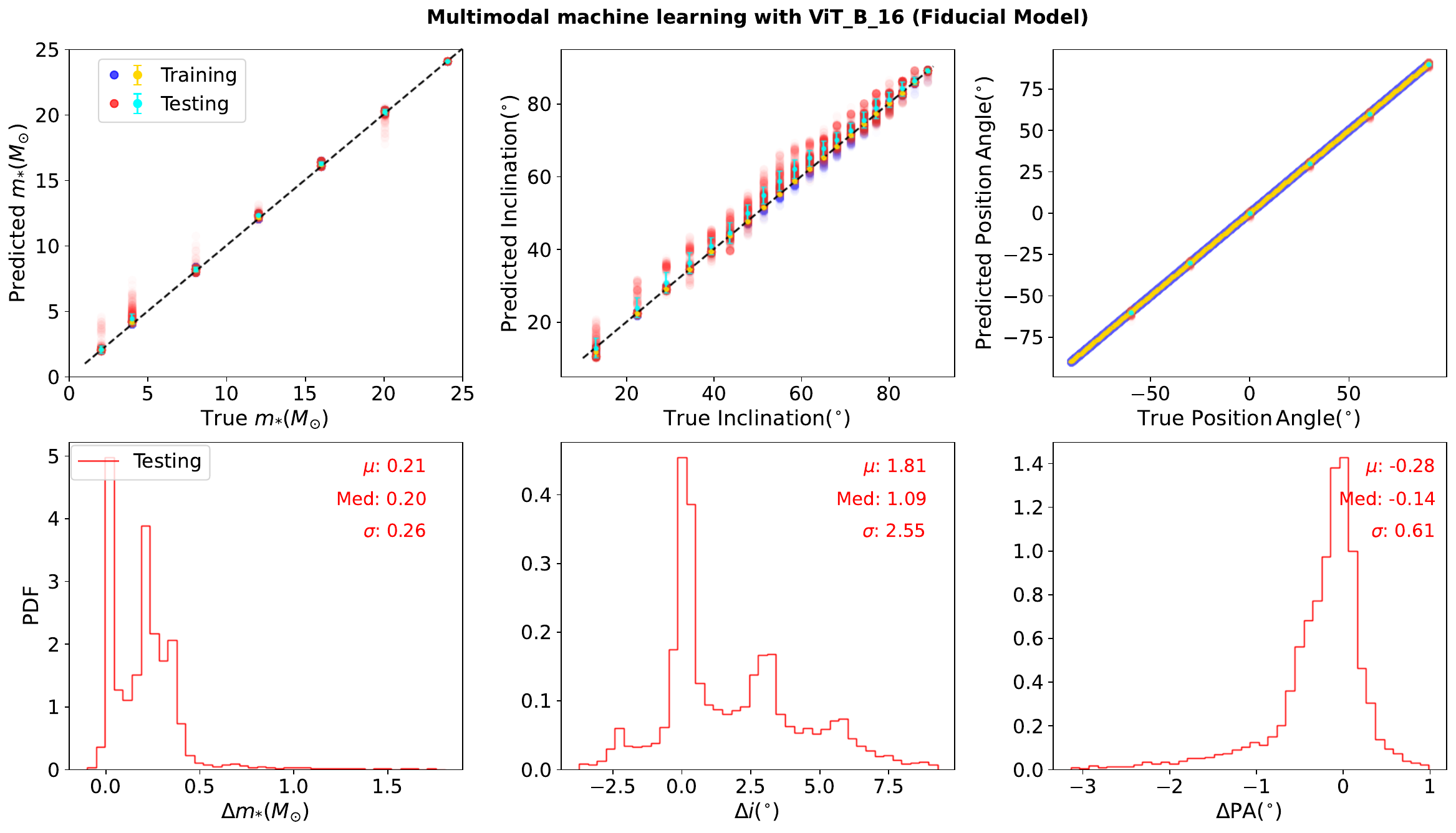}
\includegraphics[width=0.49\linewidth]{{pred_vit_l_16_uncertainty_test_0628}.pdf}
\caption{Same as Figure~\ref{fig.resnet_vit_12msun_test}, but for models trained on the full range of protostellar masses and inclination angles.}
\label{fig.resnet_vit_All_test}
\end{figure*} 

\begin{figure*}[hbt!]
\centering
\includegraphics[width=0.49\linewidth]{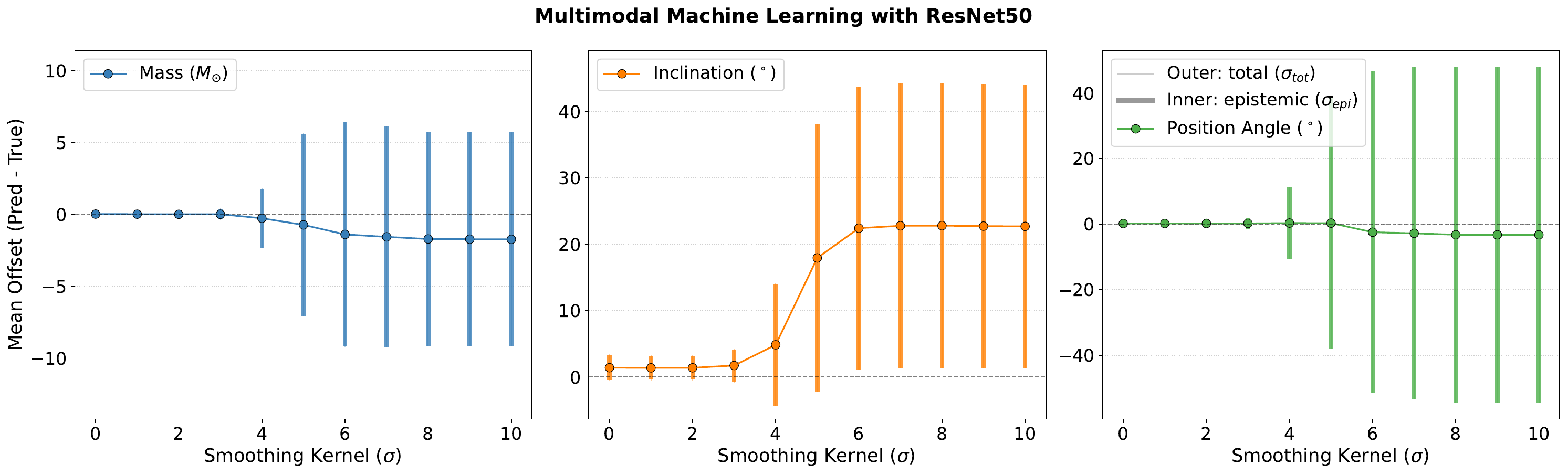}
\includegraphics[width=0.49\linewidth]{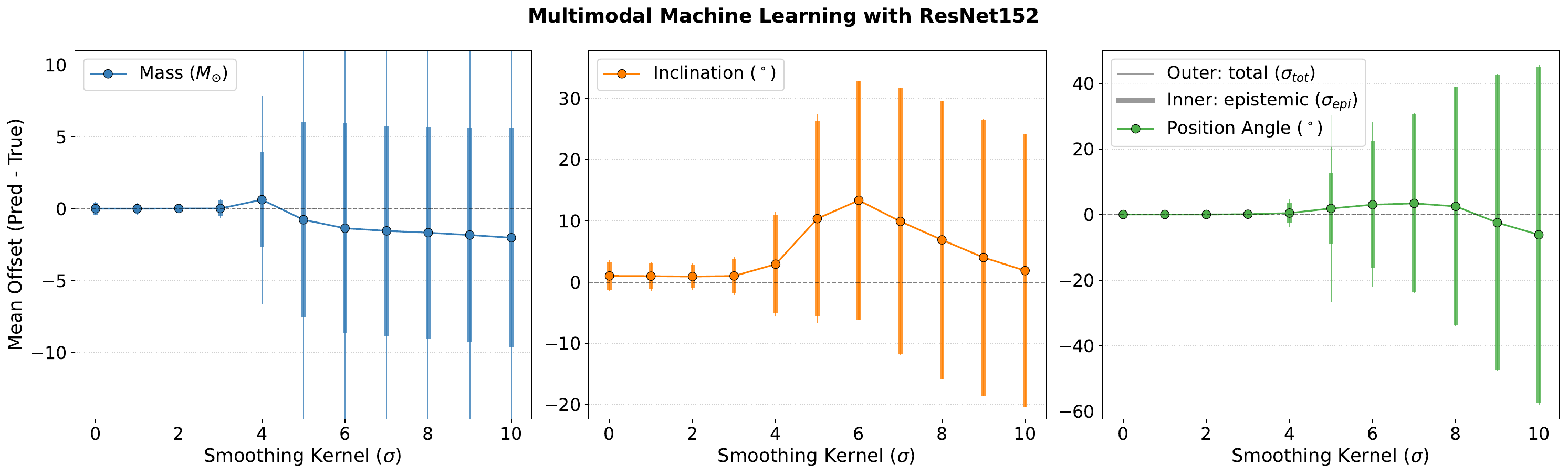}
\includegraphics[width=0.49\linewidth]{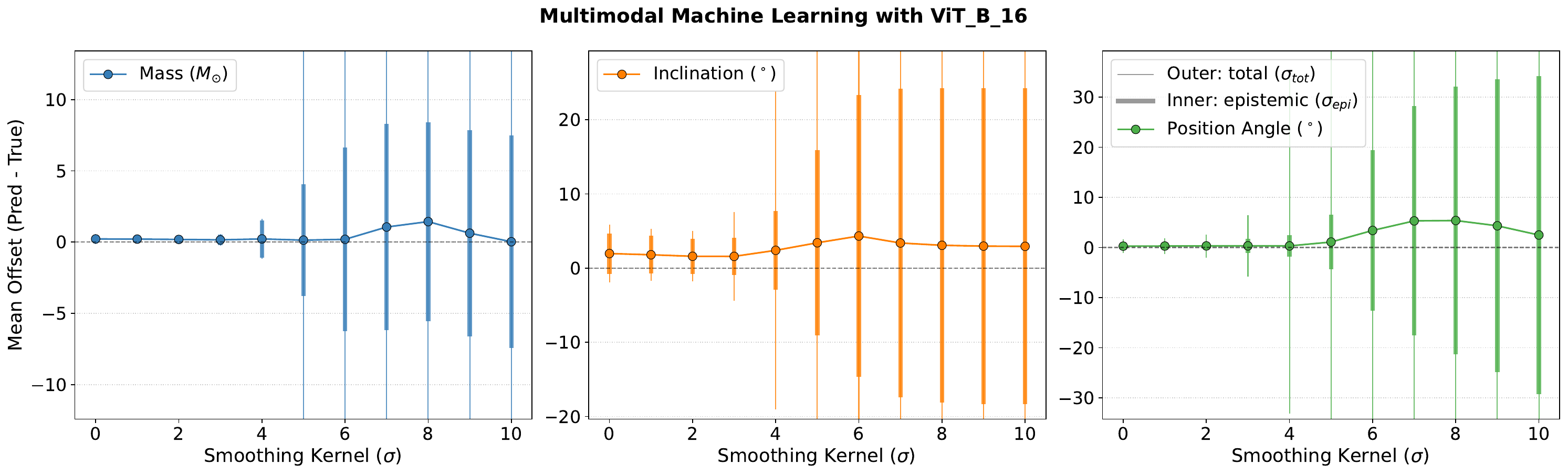}
\includegraphics[width=0.49\linewidth]{{pred_summary_smooth_vit_l_layered}.pdf}
\caption{Performance of four machine learning models on synthetic data convolved with progressively larger Gaussian kernels. From left to right, panels show results for protostellar mass, inclination angle, and position angle, including predicted means, total uncertainties, and epistemic (model) uncertainties. The total uncertainty is computed as the quadrature sum of epistemic and aleatoric (data) uncertainties. }
\label{fig.pred_summary_smooth_resnet_vit}
\end{figure*} 

\begin{figure*}[hbt!]
\centering
\includegraphics[width=0.49\linewidth]{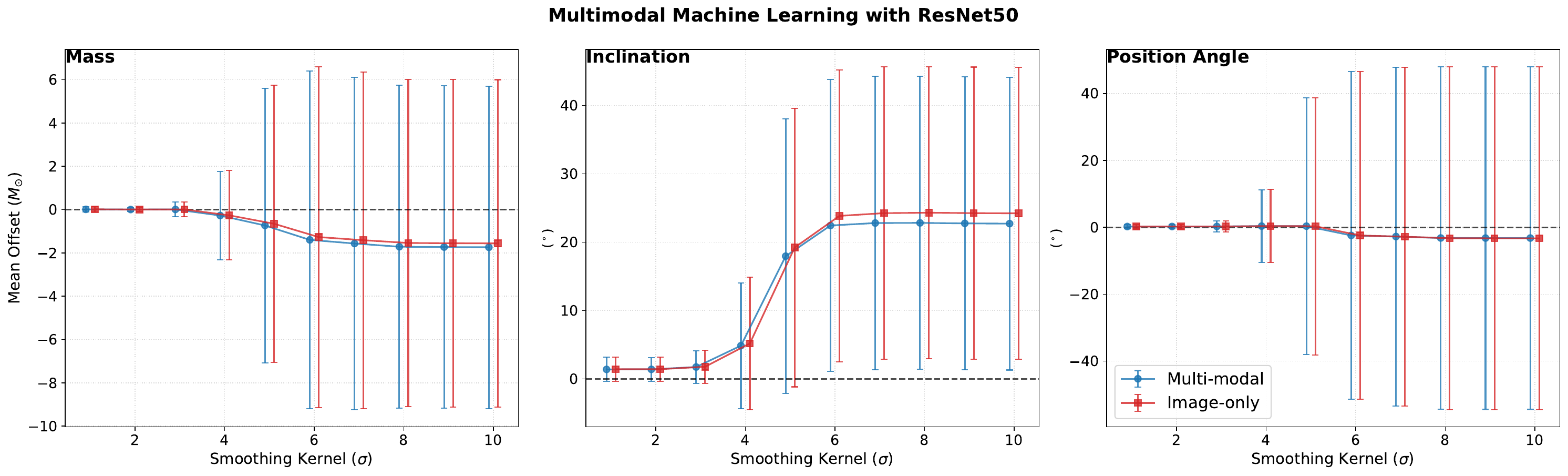}
\includegraphics[width=0.49\linewidth]{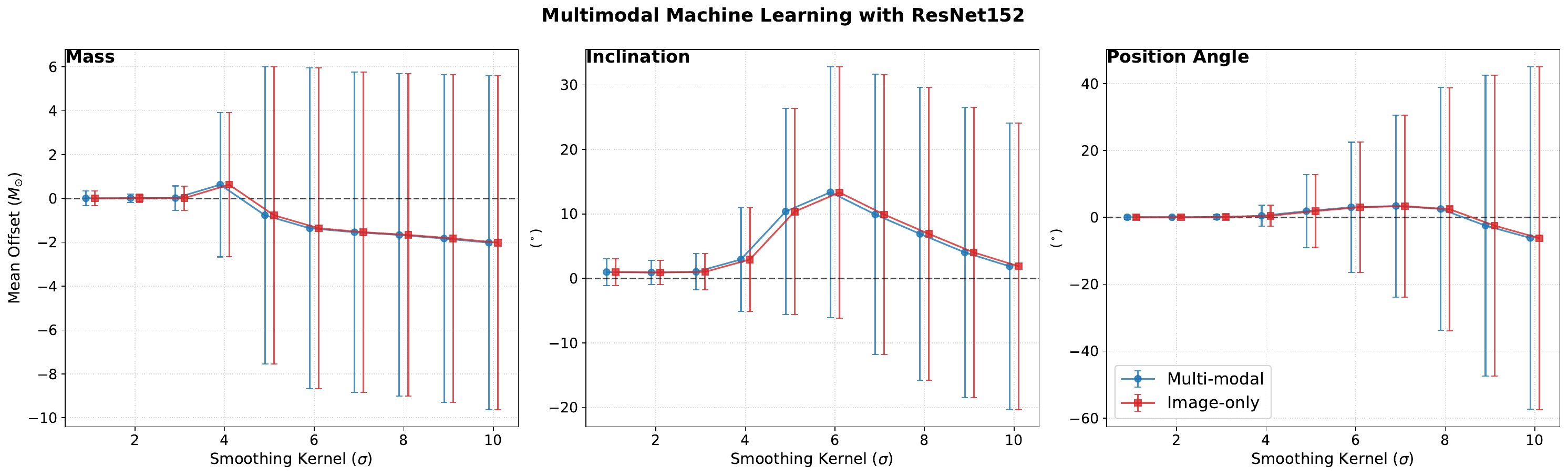}
\includegraphics[width=0.49\linewidth]{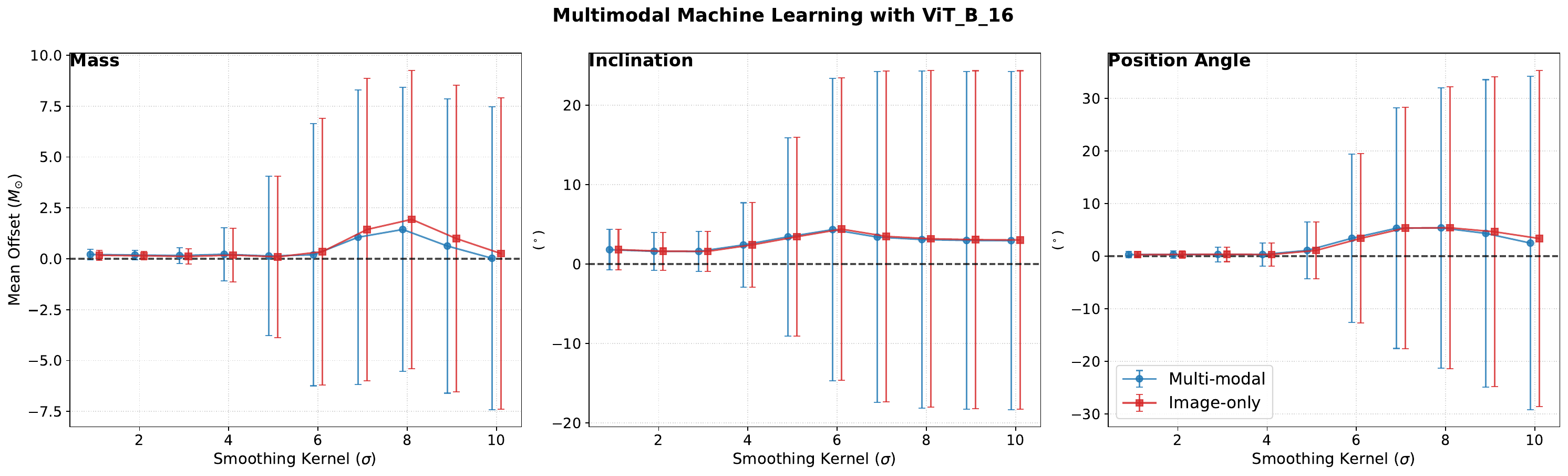}
\includegraphics[width=0.49\linewidth]{{pred_nospectrum_test_smooth_vit_l}.pdf}
\caption{Comparison of model performance for two input configurations: multi-modal (images + spectra) and image-only, evaluated on synthetic data with progressively increased Gaussian smoothing.}
\label{fig.pred_nospectrum_test_smooth_resnet_vit}
\end{figure*}

Figures~\ref{fig.resnet_vit_12msun_test}--\ref{fig.pred_nospectrum_test_smooth_resnet_vit} compare the performance of ResNet and ViT architectures. We find that ViT models generally offer superior generalization on unseen parameters and greater robustness to spatial resolution degradation (Figure~\ref{fig.pred_summary_smooth_resnet_vit}). Additionally, ablation studies (Figure~\ref{fig.pred_nospectrum_test_smooth_resnet_vit}) confirm that spatial morphology, rather than spectral kinematics, is the dominant driver of the inferred physical parameters, with spectral omission having negligible impact on accuracy for moderate smoothing.

\section{Gallery of Model Interpretability Visualizations}
\label{Gallery of Model Interpretability Visualizations}

\begin{figure*}[hbt!]
\centering
\includegraphics[width=0.49\linewidth]{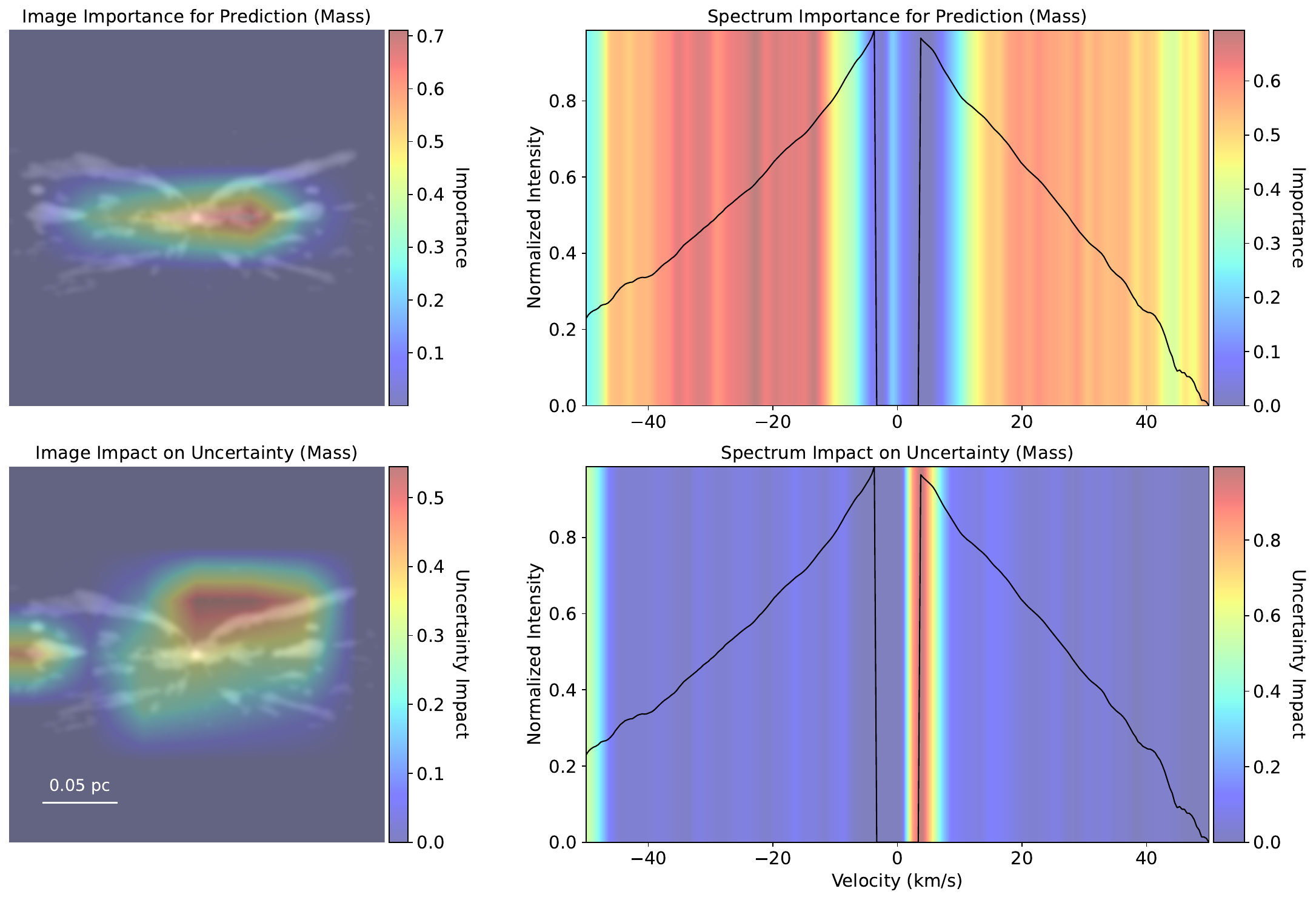}
\includegraphics[width=0.49\linewidth]{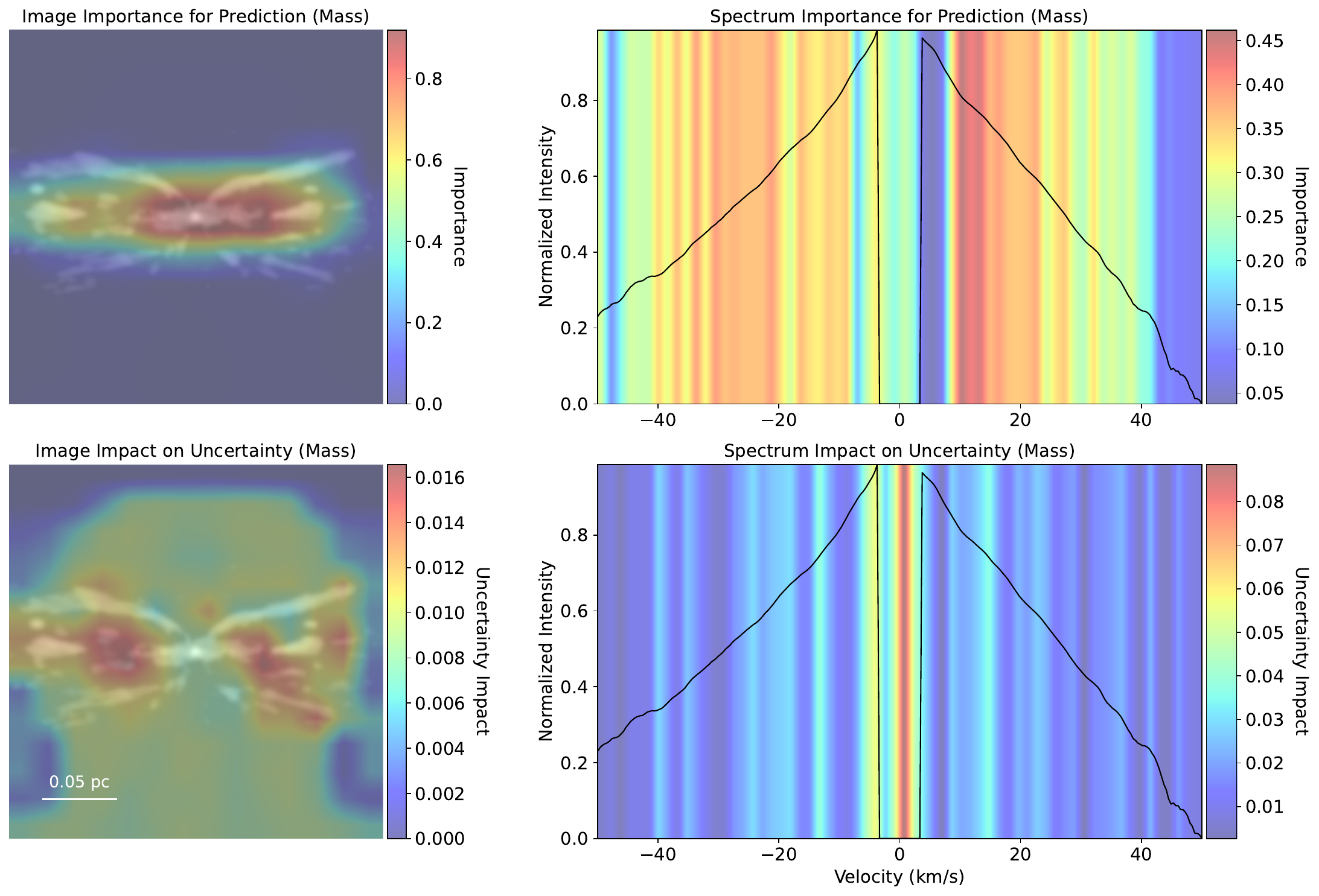}
\caption{Grad-CAM++ visualizations for outflow mass prediction and its associated uncertainty using the ResNet50 (left panels) and ViT\_L\_16 (right panels) models. In each subplot, the background shows the corresponding test outflow input, image (left) and spectrum (right). The upper row displays the attribution maps for mass prediction, while the lower row shows those for mass uncertainty prediction. The overlaid heatmaps indicate regions that contribute most strongly to the inferred quantities. }
\label{fig.GradCAMPlusPlus_resnet_vit_Smooth_0}
\end{figure*} 

\begin{figure*}[hbt!]
\centering
\includegraphics[width=0.49\linewidth]{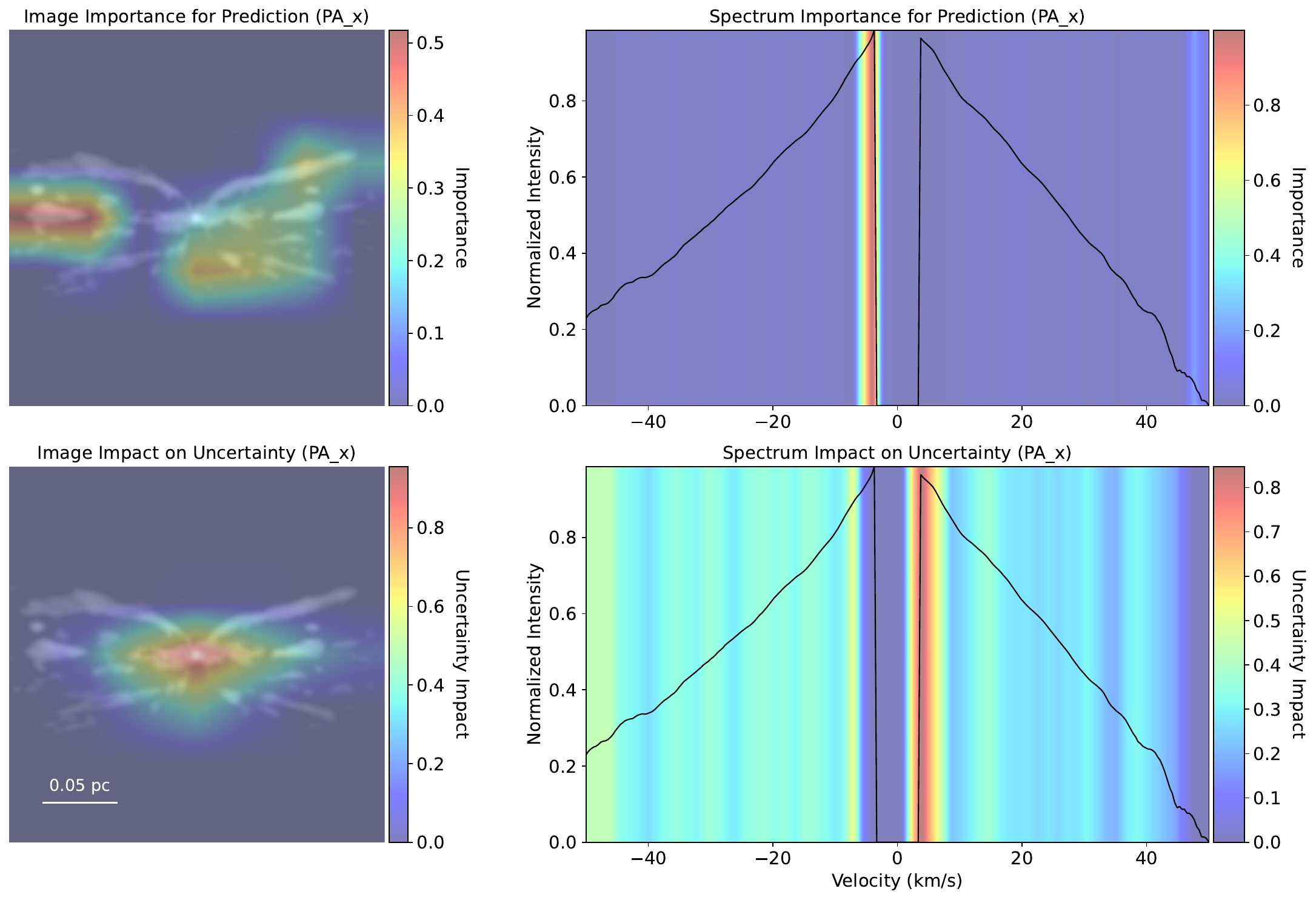}
\includegraphics[width=0.49\linewidth]{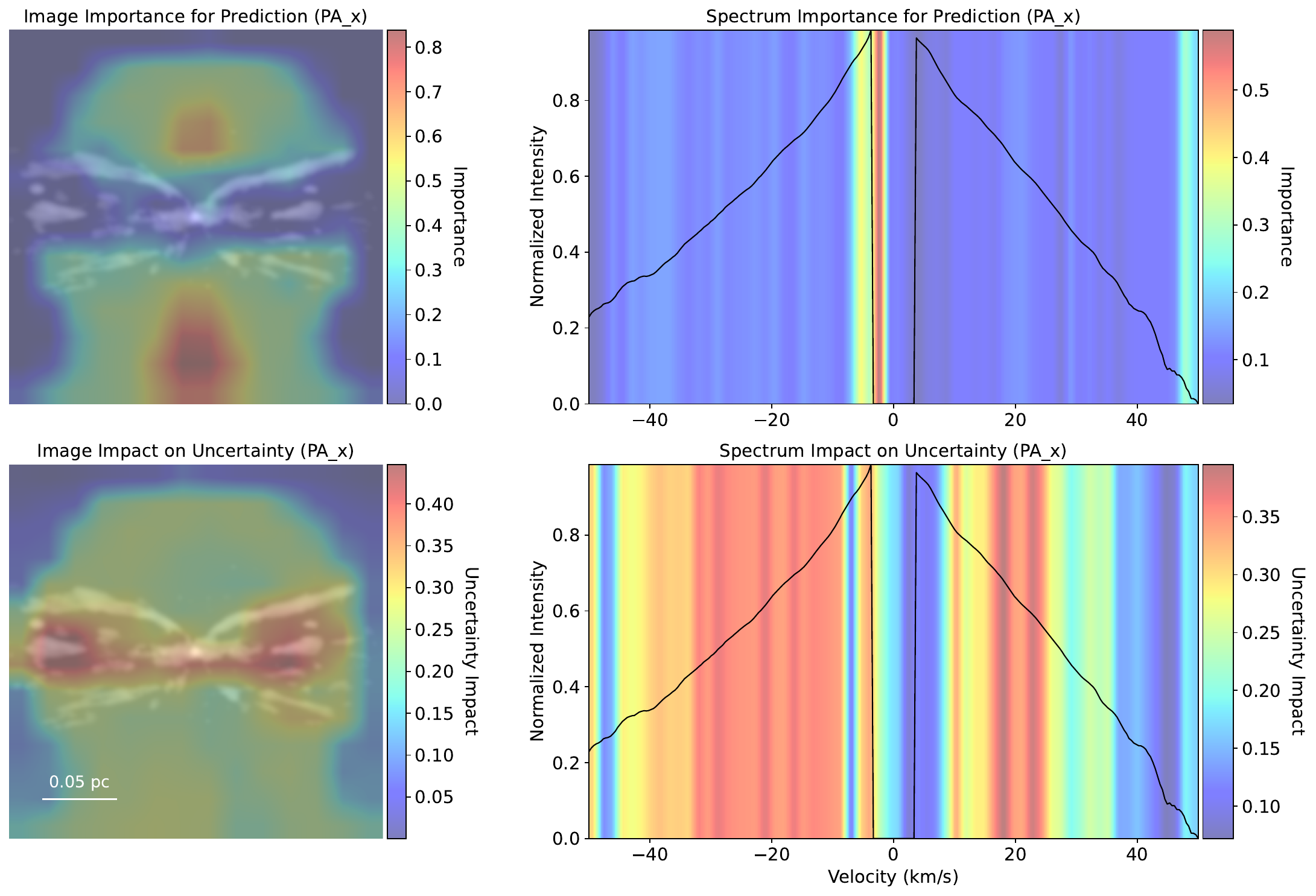}
\caption{Same as Figure~\ref{fig.GradCAMPlusPlus_resnet_vit_Smooth_0}, but for Grad-CAM++ visualizations of one component of the position angle (the cosine term) and the corresponding uncertainty prediction. }
\label{fig.GradCAMPlusPlus_resnet_vit_Smooth_2}
\end{figure*}

\begin{figure*}[hbt!]
\centering
\includegraphics[width=0.49\linewidth]{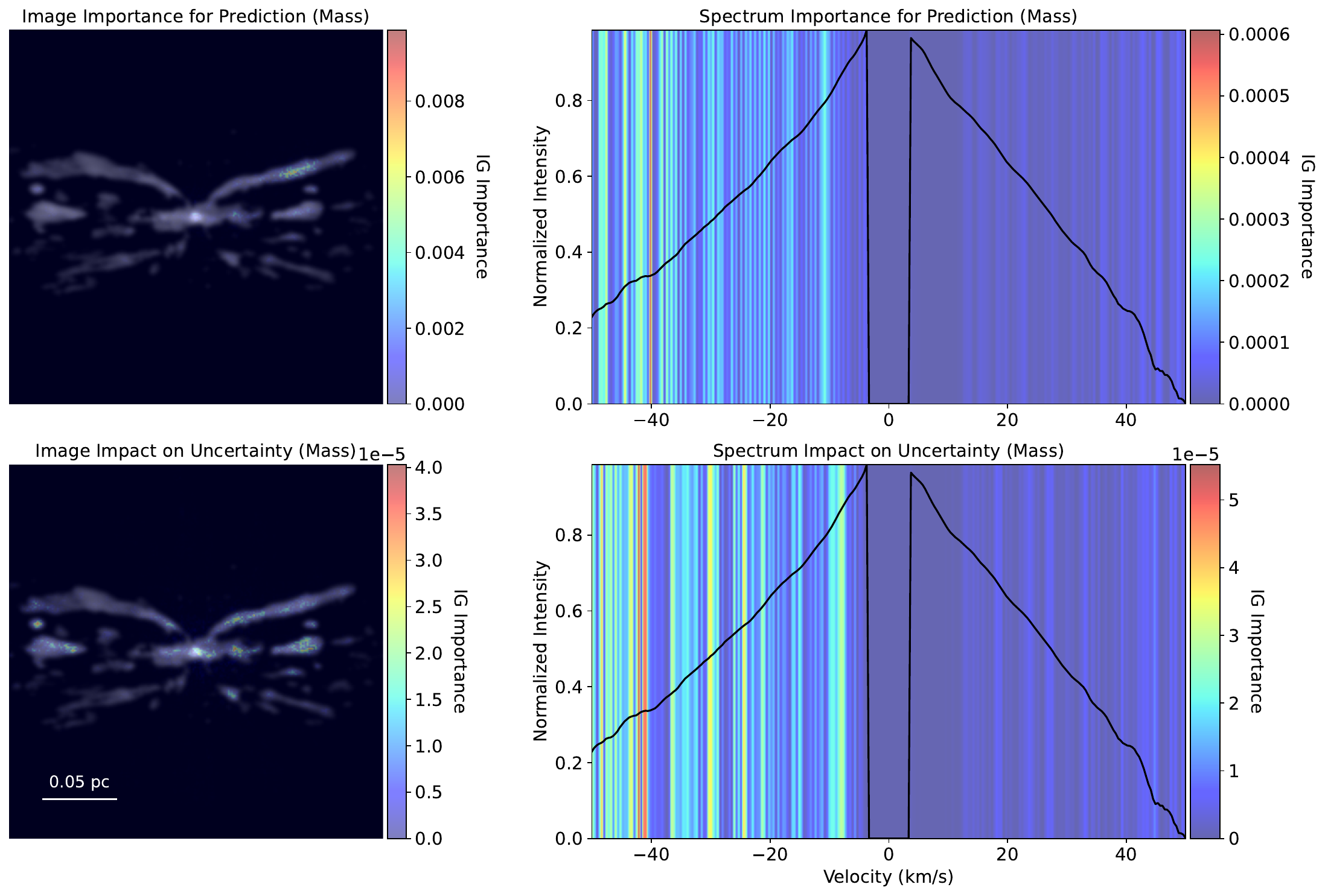}
\includegraphics[width=0.49\linewidth]{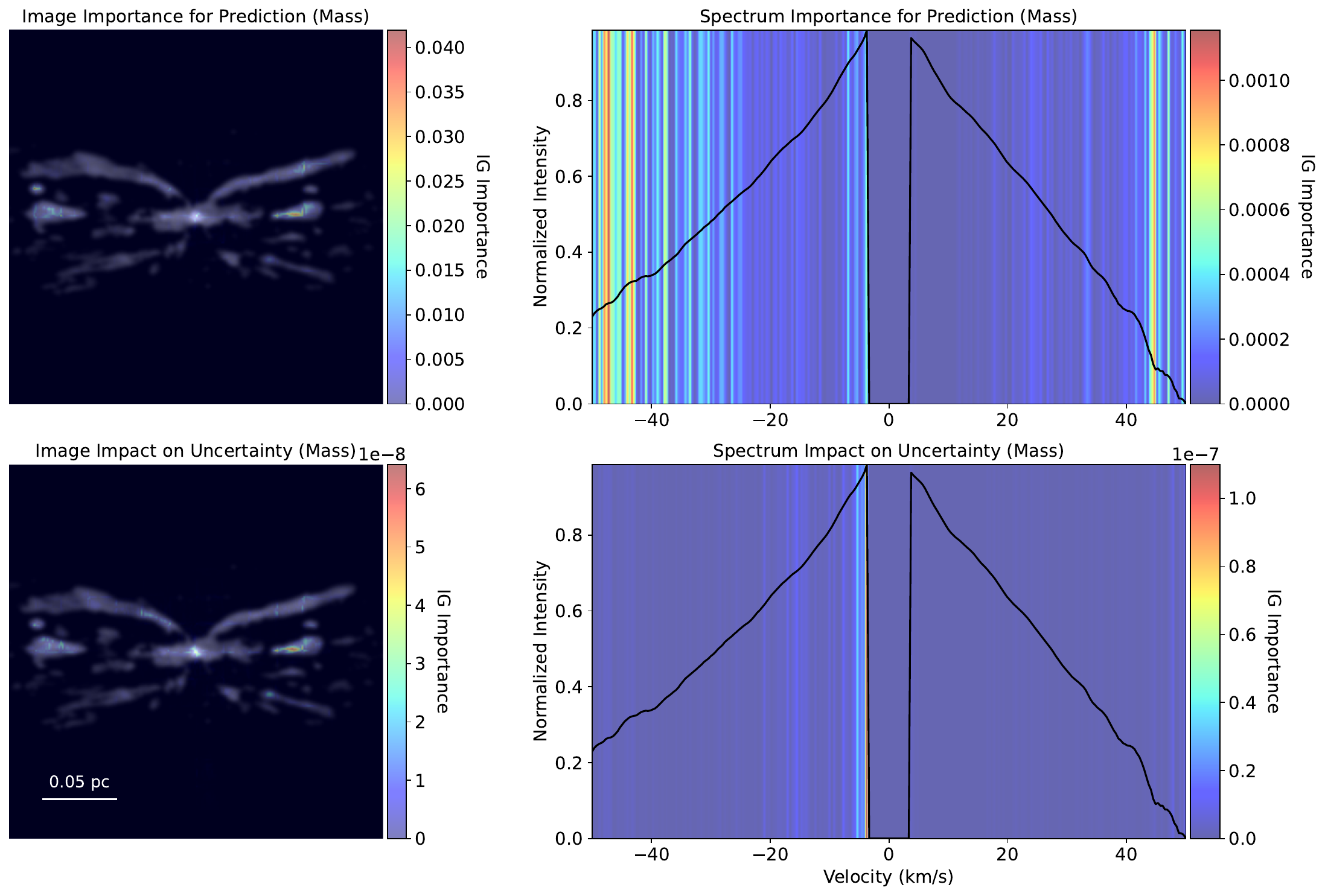}
\caption{Same as Figure~\ref{fig.GradCAMPlusPlus_resnet_vit_Smooth_0}, but showing Integrated Gradients maps for the prediction of outflow mass and its associated uncertainty using the ResNet50 (left panels) and ViT\_L\_16 (right panels) models.}
\label{fig.IG_heatmap_MANUAL_resnet_vit_mean_0}
\end{figure*} 

\begin{figure*}[hbt!]
\centering
\includegraphics[width=0.49\linewidth]{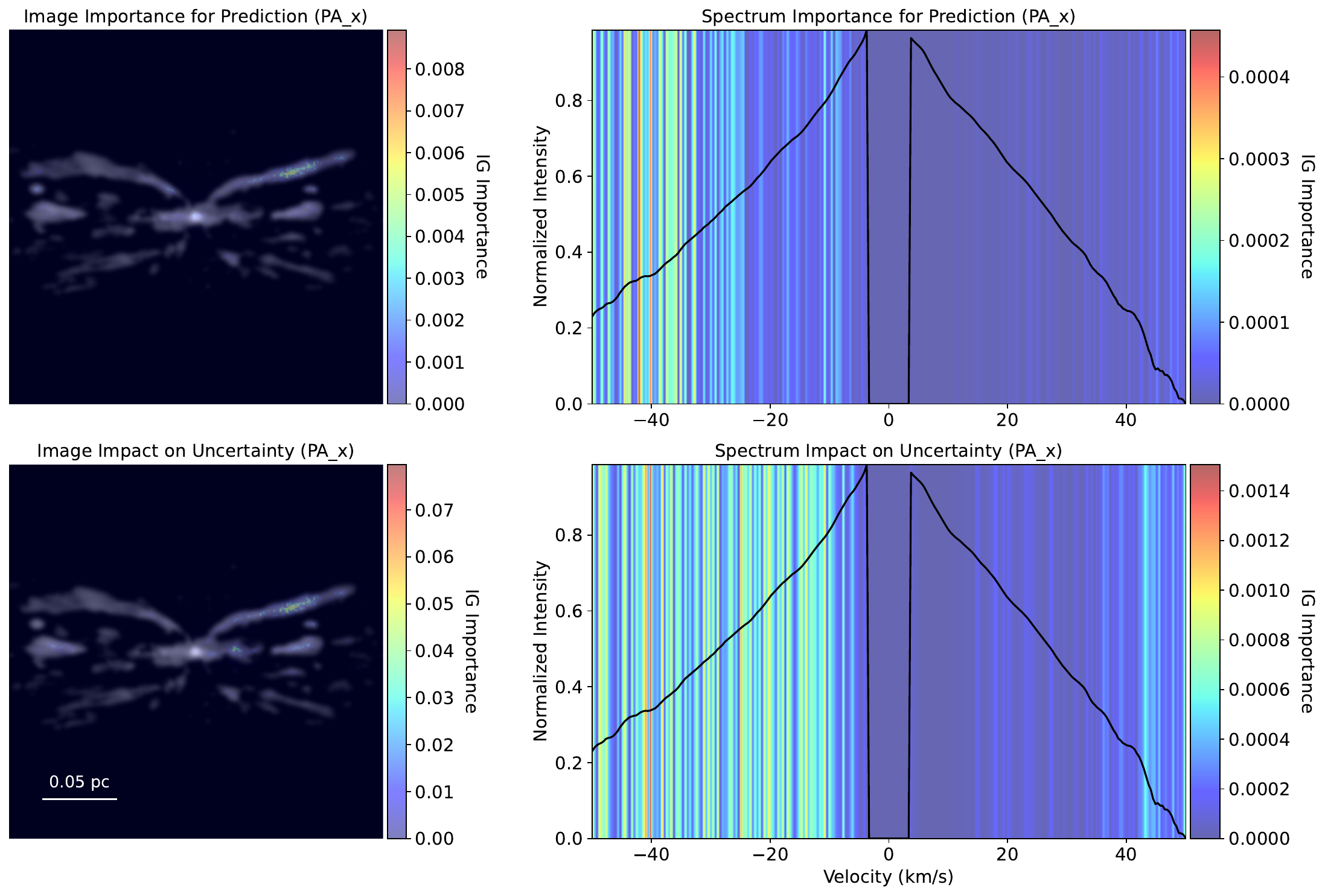}
\includegraphics[width=0.49\linewidth]{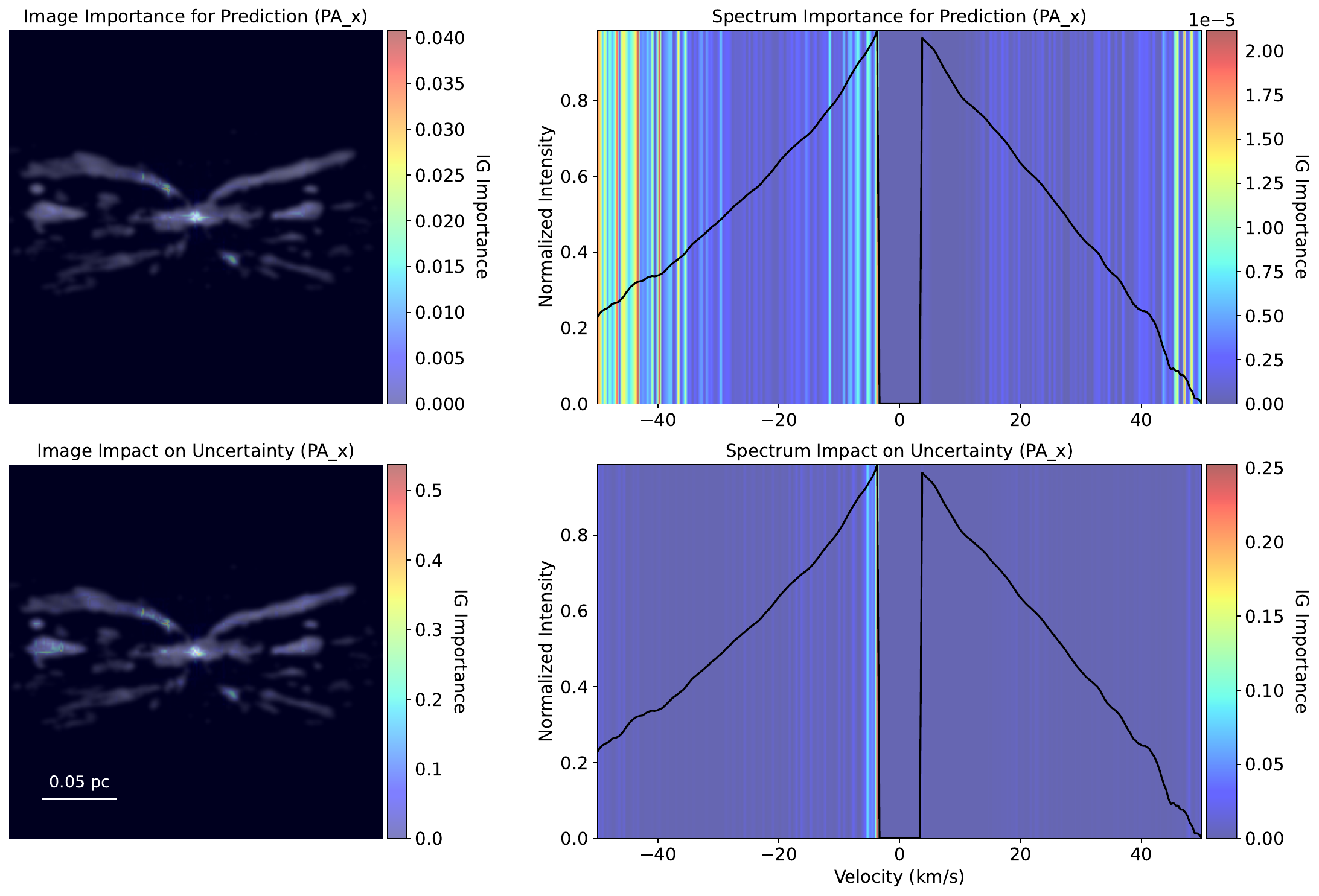}
\caption{Same as Figure~\ref{fig.IG_heatmap_MANUAL_resnet_vit_mean_0}, but showing Integrated Gradients maps for one component of the position angle (the cosine term) and its associated uncertainty.}
\label{fig.IG_heatmap_MANUAL_resnet_vit_mean_2}
\end{figure*} 

\begin{figure*}[hbt!]
\centering
\includegraphics[width=0.49\linewidth]{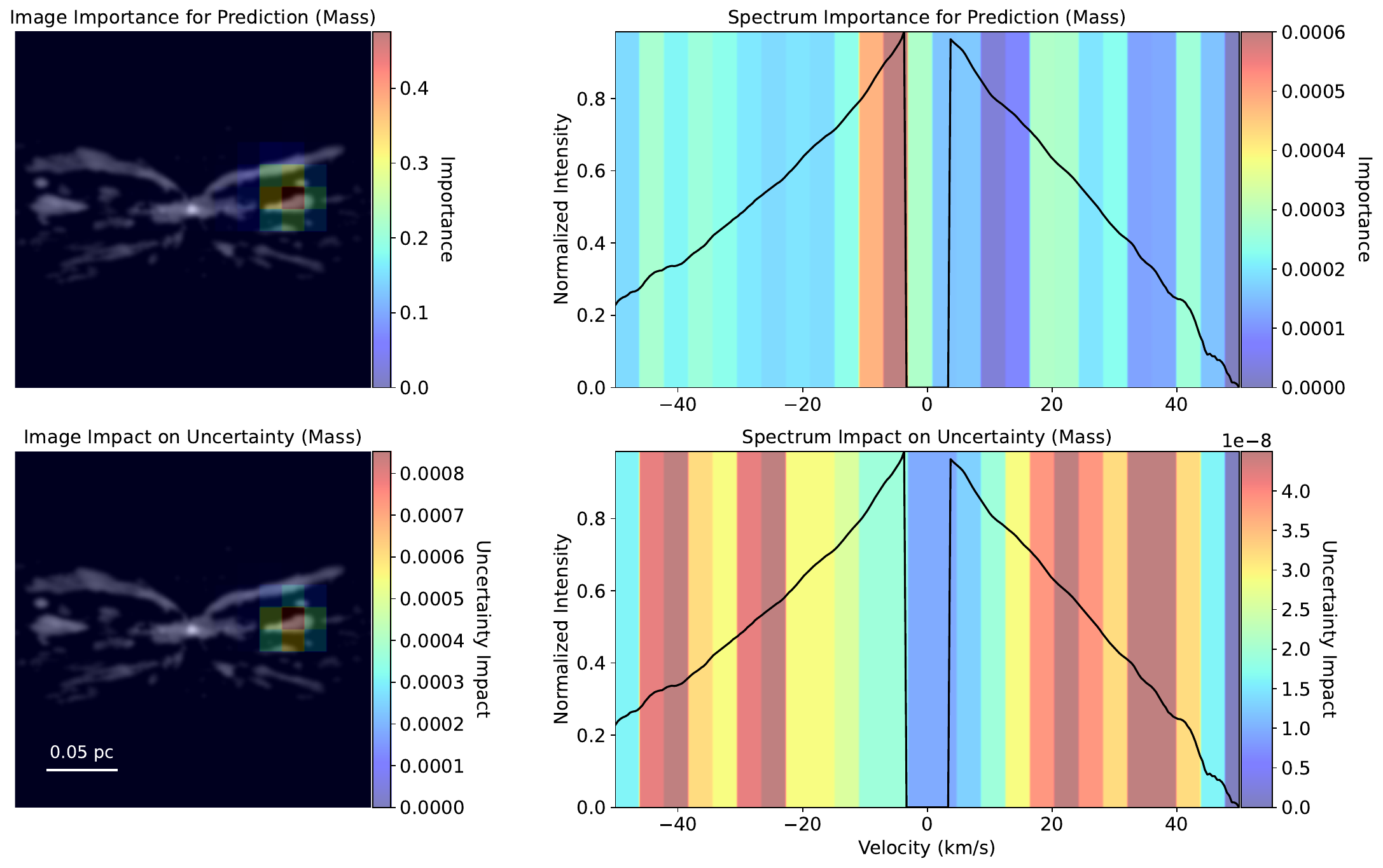}
\includegraphics[width=0.49\linewidth]{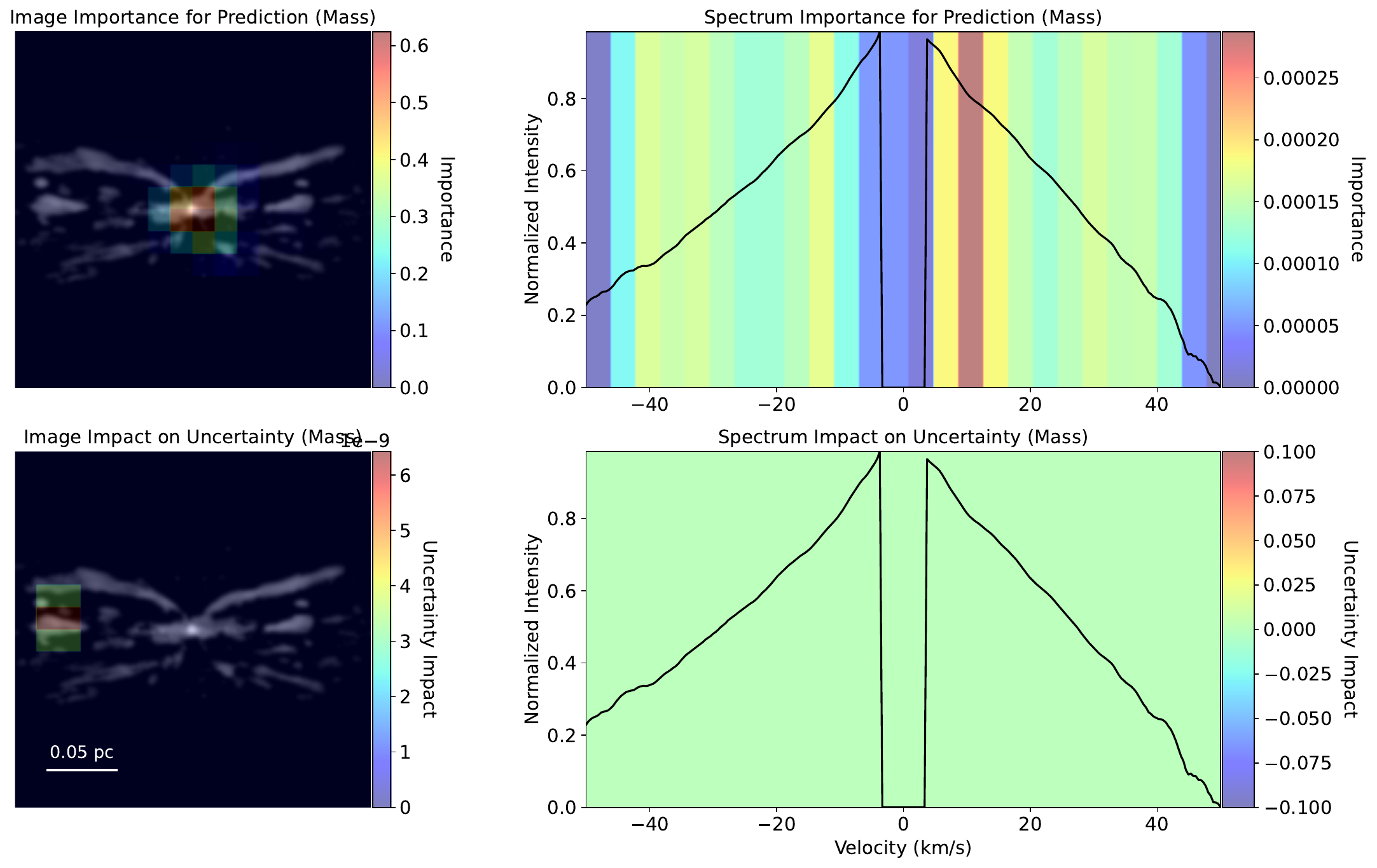}
\caption{Same as Figure~\ref{fig.GradCAMPlusPlus_resnet_vit_Smooth_0}, but showing Occlusion Sensitivity Analysis results for outflow mass prediction and its associated uncertainty using the ResNet50 (left panels) and ViT\_L\_16 (right panels) models.}
\label{fig.occlusion_uncertainty_resnet_vit_output_0}
\end{figure*} 

\begin{figure*}[hbt!]
\centering
\includegraphics[width=0.49\linewidth]{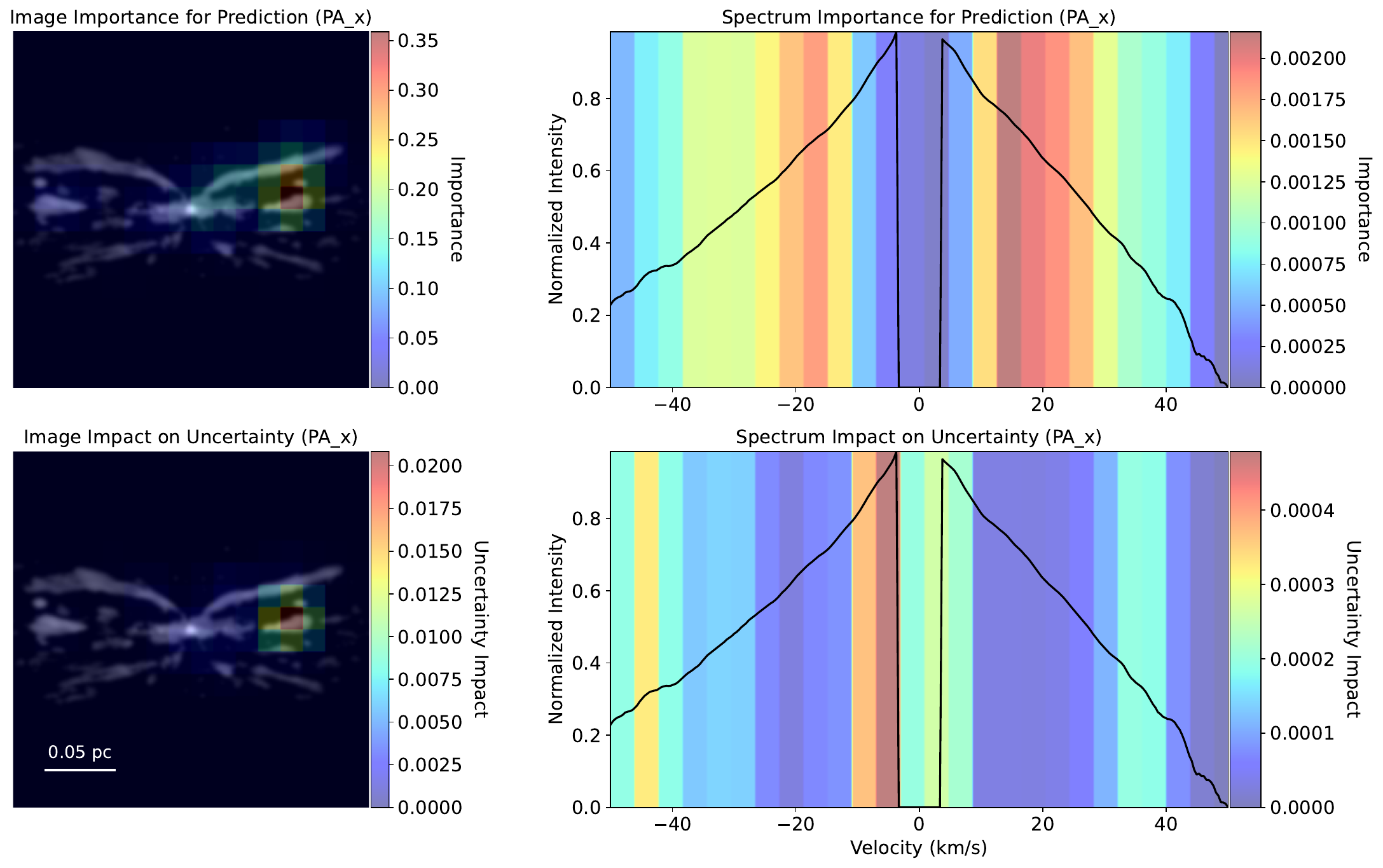}
\includegraphics[width=0.49\linewidth]{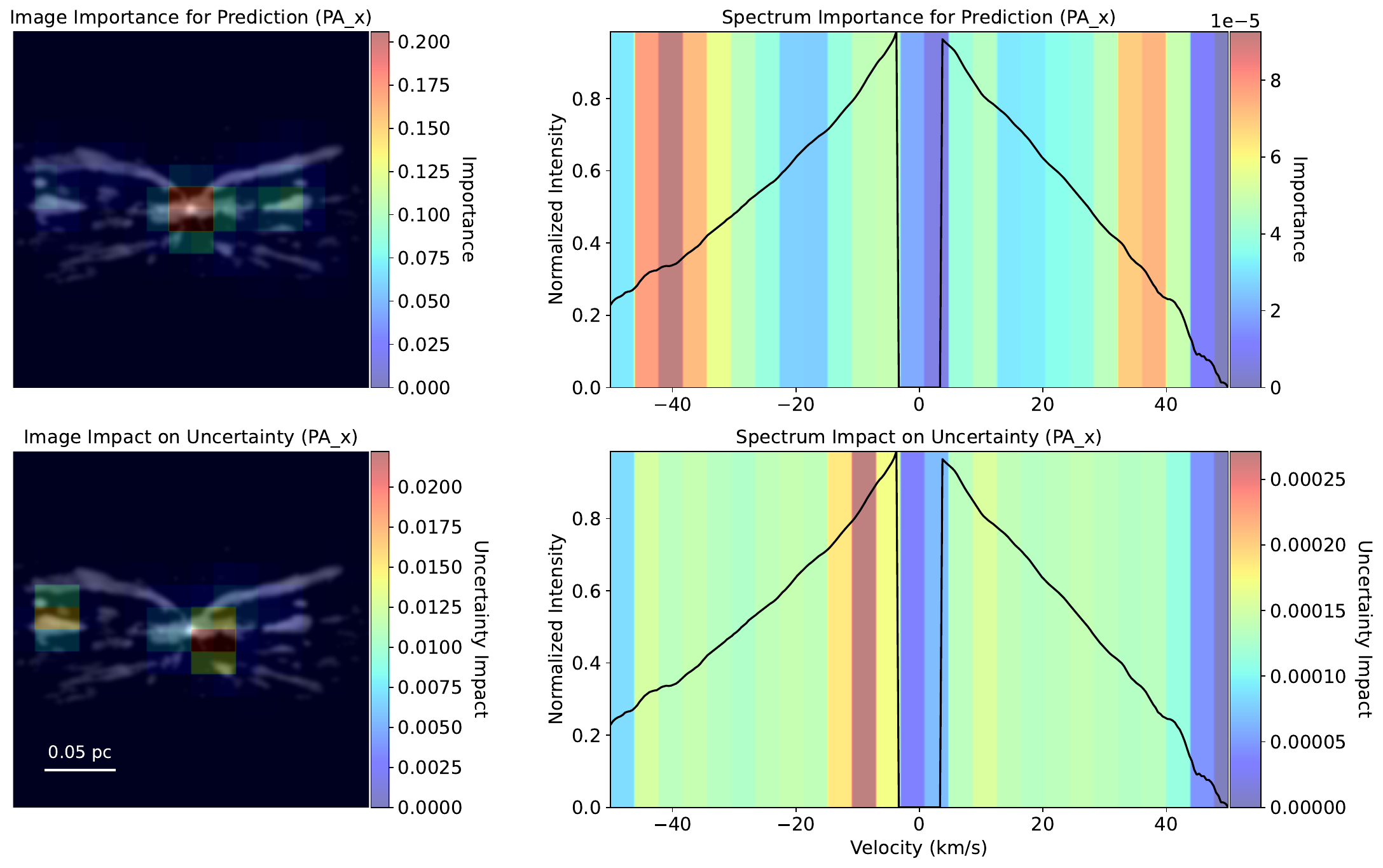}
\caption{Same as Figure~\ref{fig.occlusion_uncertainty_resnet_vit_output_0}, but showing Occlusion Sensitivity Analysis results for one component of the position angle (the cosine term) and its associated uncertainty.}
\label{fig.occlusion_uncertainty_resnet_vit_output_2}
\end{figure*} 

In this section, we present a comprehensive gallery of interpretability visualizations comparing the decision-making processes of the ResNet50 and ViT\_L\_16 architectures. We display Smooth Grad-CAM++ attribution maps for predicted outflow mass and the cosine component of the position angle, along with their respective uncertainty estimates, in Figures~\ref{fig.GradCAMPlusPlus_resnet_vit_Smooth_0} and \ref{fig.GradCAMPlusPlus_resnet_vit_Smooth_2}. Complementing these, we provide pixel-level attribution via Integrated Gradients for the same parameters in Figures~\ref{fig.IG_heatmap_MANUAL_resnet_vit_mean_0} and \ref{fig.IG_heatmap_MANUAL_resnet_vit_mean_2}. Finally, the results of the Occlusion Sensitivity Analysis are presented in Figures~\ref{fig.occlusion_uncertainty_resnet_vit_output_0} and \ref{fig.occlusion_uncertainty_resnet_vit_output_2}, verifying the causal importance of the identified spatial features.

Collectively, these complementary visualizations confirm that the models drive their inference from physically meaningful spatial structures, such as cavity walls and terminal lobes, providing robust evidence that the learned representations are grounded in the actual physical morphology of the outflows rather than observational artifacts.

\end{CJK*}

\end{document}